# Toward a Sustainable Low-Altitude Economy: A Survey of Energy-Efficient RIS-UAV Networks

Manzoor Ahmed, Aized Amin Soofi, Feroz Khan, Salman Raza, Wali Ullah Khan, Lina Su, Fang Xu, and Zhu Han, *Fellow, IEEE,*

***Abstract*—The integration of reconfigurable intelligent surfaces (RIS) into unmanned aerial vehicle (UAV) networks presents a transformative solution for achieving energy-efficient and reliable communication, particularly within the rapidly expanding low-altitude economy (LAE). As UAVs facilitate diverse aerial services—spanning logistics to smart surveillance—their limited energy reserves create significant challenges. RIS effectively addresses this issue by dynamically shaping the wireless environment to enhance signal quality, blackuce power consumption, and extend UAV operation time, thus enabling sustainable and scalable deployment across various LAE applications. This survey provides a comprehensive review of RIS-assisted UAV networks, focusing on energy-efficient design within LAE applications. We begin by introducing the fundamentals of RIS, covering its operational modes, deployment architectures, and roles in both terrestrial and aerial environments. Next, advanced energy efficiency (EE)-driven strategies for integrating RIS and UAVs. Techniques such as trajectory optimization, power control, beamforming, and dynamic resource management are examined. Emphasis is placed on collaborative solutions that incorporate UAV-mounted RIS, wireless energy harvesting (EH), and intelligent scheduling frameworks. We further categorize RIS-enabled schemes based on key performance objectives relevant to LAE scenarios. These objectives include sum rate maximization, coverage extension, quality of service (QoS) guarantees, secrecy rate improvement, latency blackuction, and age of information (AoI) minimization. The survey also delves into RIS-UAV synergy with emerging technologies like multi-access edge computing (MEC), non-orthogonal multiple access (NOMA), vehicle-to-everything (V2X) communication, and wireless power transfer (WPT). Finally, we outline open research challenges and future directions, emphasizing the critical role of energy-aware, RIS-enhanced UAV networks in shaping scalable, sustainable, and intelligent infrastructures within the LAE.**

Manzoor Ahmed, Lina Su, and Fang Xu are with the School of Computer and Information Science and also with Institute for AI Industrial Technology Research, Hubei Engineering University, Xiaogan City, 432000, China (e-mails: manzoor.achakzai@gmail.com, lina.su@whu.edu.cn, xf@hbeu.edu.cn), Aized Amin Soofi is with the Department of Computer Science, National University of Modern Languages Faisalabad, 38000, Pakistan (e-mail:aizedamin@gmail.com ).

Feroz Khan is with the School of Electronic Engineering, Beijing University of Posts and Telecommunications, Beijing, China (e-mail: ferozkhan687@gmail.com).

Salman Raza is with the Department of Computer Science, National Textile University Faisalabad, 38000, Pakistan (e-mail: salmanraza@ntu.edu.pk).

Wali Ullah Khan is with the School of Computing, Gachon University, Seongnam-si 13120, Republic of Korea (e-mails: waliullahkhan30@gmail.com).

Zhu Han is with the Department of Electrical and Computer Engineering at the University of Houston, Houston, TX 77004 USA, and also with the Department of Computer Science and Engineering, Kyung Hee University, Seoul, South Korea, 446-701 (e-mail: hanzhu22@gmail.com).

*(Corresponding Authors: Manzoor Ahmed and Wali Ullah Khan)*

## I. Introduction

THE low-altitude economy (LAE) is emerging as a transformative domain, encompassing a broad range of applications such as aerial logistics, smart surveillance, urban air mobility, and public safety [1]. It leverages the capabilities of manned and unmanned aerial vehicles (UAVs) to deliver efficient, responsive services in near-ground airspace [2], [3]. Realizing the potential of LAE hinges on energy-efficient, scalable, and intelligent wireless communication infrastructure that can operate reliably in dense, low-altitude environments.

As wireless communications advance, the integration of UAVs with reconfigurable intelligent surfaces (RIS) emerges as a promising solution for enhancing energy efficiency (EE) and communication performance. With the transition towards sixth-generation (6G) networks, there is a heightened focus on achieving ultra-high data rates, low latency, and minimal energy consumption [4], [5]. Among the various technologies being exploblack, RIS has garneblack substantial attention due to its capability to dynamically manipulate the propagation environment [6].

RIS technology represents a significant leap forward by introducing an intelligent layer that enhances signal propagation through controlled reflection and phase adjustments [7], [8]. In contrast to traditional systems that heavily depend on transmitter and receiver efficiencies [9], RIS operates in passive, active, and hybrid modes, showcasing its versatility. These surfaces provide substantial enhancements in communication coverage and reliability, making them suitable for diverse environmental conditions [10]. Moreover, RIS's ability to navigate around physical obstructions and minimize signal loss elevates its status as a pivotal technology for advancing connectivity and improving the user experience [11]. This unique capability further underscores its importance in modern communication systems [12].

UAVs have evolved into essential components of modern wireless communication systems, largely due to their high mobility, capability to establish line-of-sight (LoS) communication links, and cost-effectiveness [13]. These systems are applied across a wide range of scenarios that demand enhanced communication quality, such as disaster management, surveillance, and dynamic communication networks [14]. Furthermore, optimizing UAV trajectories while also addressing resource allocation can lead to substantial improvements in critical metrics like throughput, secrecy rate, and EE. In the context of LAE, these enhancements are crucial for ensuring that UAVs operate efficiently across diverse urban and remote

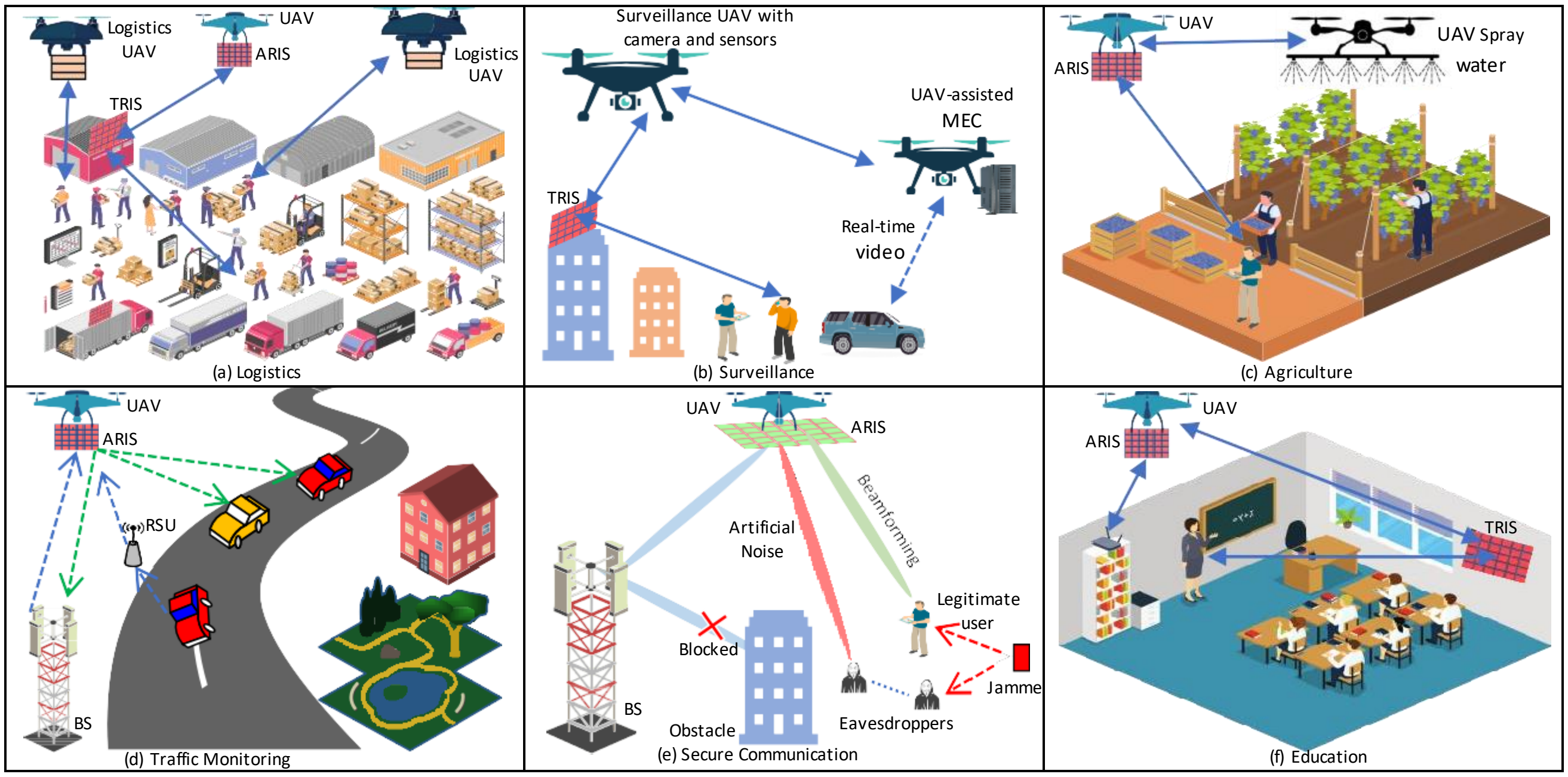

Fig. 1: Illustration of various real-life energy-saving design scenarios enabled by low-altitude UAV-RIS technology, including logistics, surveillance, smart agriculture, traffic monitoring, secure communication, and education, showcasing the potential of intelligent aerial communication systems in the LAE.

TABLE I: List of Important Abbreviations

| Abbreviation | Definition |
|---|---|
| AoI | Age of Information |
| AP | Access Point |
| ARIS | Aerial Reconfigurable Intelligent Surface |
| BCD | Block Coordinate Descent |
| BS | Base Station |
| CSI | Channel State Information |
| DDPG | Deep Deterministic Policy Gradient |
| DRL | Deep Reinforcement Learning |
| EE | Energy Efficiency |
| EH | Energy Harvesting |
| ES | Edge Server |
| GT | Ground Terminal |
| GUE | Ground User Equipment |
| IoT | Internet of Things |
| LAE | Low-altitude Economy |
| LoS | Line of Sight |
| MEC | Multi-access Edge Computing |
| ML | Machine Learning |
| mmWave | millimeter Wave |
| NLoS | Non-Line-of-Sight |
| NOMA | Non-Orthogonal Multiple Access |
| NTNs | Non Terrestrial Networks |
| PLS | Physical Layer Security |
| PPO | Proximal Policy Optimization |
| QoS | Quality of Service |
| RISs | Reconfigurable Intelligent Surfaces |
| SAC | Soft Actor-Critic |
| SCA | Successive Convex Approximation |
| SCE | Secure Computation Efficiency |
| SDR | Semidefinite Relaxation |
| SNR | Signal to Noise Ratio |
| SOP | Secrecy Outage Probability |
| STAR-RIS | Simultaneously Transmitting and Reflecting RIS |
| THz | TeraHertz |
| TRIS | Terrestrial Reconfigurable Intelligent Surface |
| UAVs | Unmanned Aerial Vehicles |
| UE | User Equipment |
| WPCN | Wireless Poweblack Communication Network |
| WPT | Wireless Power Transfer |

use cases, depicted in Fig. 1.

In urban environments, maintaining consistent LoS communication links between UAVs and ground users presents significant challenges, primarily due to various obstructions such as buildings [15]. A promising solution to this issue lies in integrating RIS with UAV communication systems. Specifically, a RIS-assisted UAV communication system leverages the mobility of UAVs alongside the advanced signal manipulation capabilities of RIS [16]. This synergy not only enhances performance but also blackuces energy consumption. Within the LAE framework, this integration proves particularly advantageous, enabling scalable aerial services without increasing operational costs or energy demands. By allowing UAVs to communicate effectively with ground users along planned trajectories, signals are reflected and optimized by RIS, resulting in substantial improvements in connectivity.

This survey provides a comprehensive review of recent advancements and future challenges in RIS-enabled UAV communications, with a particular emphasis on EE and its implications for 6G applications. We will explore the fundamental principles of RIS technology and its seamless integration with UAVs, alongside potential applications across various communication scenarios. Moreover, we will discuss optimization techniques related to UAV trajectories and RIS configurations, highlighting both the benefits and limitations of current methodologies. By examining these aspects within the context of LAE, we aim to offer valuable insights into the development of robust, energy-efficient wireless networks that harness the synergy between UAVs and RIS technology.

### *A. Related Surveys*

The joint application of RIS and UAVs has recently emerged as a promising direction for next-generation wireless networks.

Several surveys have been published on this topic, yet their coverage remains fragmented in terms of scope, performance benchmarking, and integration with enabling technologies. Table II provides a structublack comparison of existing RIS–UAV surveys, highlighting their focus areas and limitations.

Early surveys, such as Pogaku et al. [17] and Park et al. [18], initiated the discussion on RIS-assisted UAV communication but remained preliminary, offering high-level frameworks without detailed analysis of EE. Rahmatov et al. [19] expanded the perspective to 6G visions but did not address EE trade-offs or LAE constraints. More recent contributions have broadened the scope: Ahmed et al. [20] consideblack RIS–UAV–MEC integration, Lee et al. [21] exploblack RIS–UAV systems in the context of ISAC and multiple access, Kumar et al. [22] reviewed RIS-assisted THz UAV communications, and Ahmed et al. [23] analyzed UAV–RIS networks from a physical layer security perspective. In particular, Ning et al. [24] provide the most comprehensive survey to date on UAV–RIS integration in 6G. Their work addresses architectural aspects, applications, and design challenges, but it does not deliver an EE-centric benchmarking framework or performance trade-off analysis in LAEs.

In contrast, our work offers a unified and in-depth survey of RIS–UAV networks with a dedicated emphasis on EE in LAEs. Beyond reviewing fundamentals, it systematically evaluates performance trade-offs across sum rate, secrecy, latency, and overall system performance. Furthermore, it investigates how RIS–UAV systems can be jointly optimized with enabling technologies such as multi-access edge computing (MEC), non-orthogonal multiple access (NOMA), vehicle-to-everything (V2X), and wireless power transfer (WPT), offering a holistic view that is absent from prior works. This positions our survey as a distinctive contribution in bridging RIS–UAV research with sustainable and energy-aware design objectives.

### B. Motivation and Contribution

The integration of RIS into UAV-assisted networks significantly impacts the LAE—a rapidly evolving domain that includes urban air mobility, aerial logistics, remote sensing, and emergency response. These LAE applications necessitate energy-efficient, adaptive, and high-performance aerial communication infrastructures. However, UAVs encounter inherent limitations regarding energy supply, link reliability, and real-time adaptability. RIS emerges as a transformative solution by dynamically reconfiguring the wireless propagation environment, which enhances signal quality, blackuces power consumption, and improves EE. This capability is crucial for developing sustainable and resilient communication in low-altitude, high-density airspace. As a result, there is an increasing demand for a thorough investigation into how RIS can be utilized to tackle the communication and energy challenges specific to LAE-oriented UAV networks.

- This survey presents a comprehensive analysis of RIS-assisted UAV networks with a focus on energy-efficient designs for LAE applications. It begins by outlining the fundamentals of RIS, its operation modes, and deployment strategies in both terrestrial and aerial environments. The survey further explores key energy-saving strategies such as trajectory optimization, power control, and beamforming, alongside collaborative methods including UAV-mounted RIS deployment, energy harvesting, scheduling, and resource allocation.
- Systematic classification of RIS-based approaches that optimize EE while maintaining communication performance in LAE scenarios. It discusses methods aimed at maximizing sum rate, expanding coverage, ensuring quality of service (QoS), improving secrecy rate, blackucing latency, and minimizing age of information (AoI), which quantifies information freshness as the time elapsed since the last received update. The study also examines the integration of RIS-UAV networks with emerging technologies. Summary tables are provided to compare methodologies, performance metrics, and research objectives across existing works.
- Lastly, this survey identifies key challenges and future research directions critical to developing energy-aware, scalable, and intelligent RIS-enhanced UAV systems for sustainable operation within the LAE.

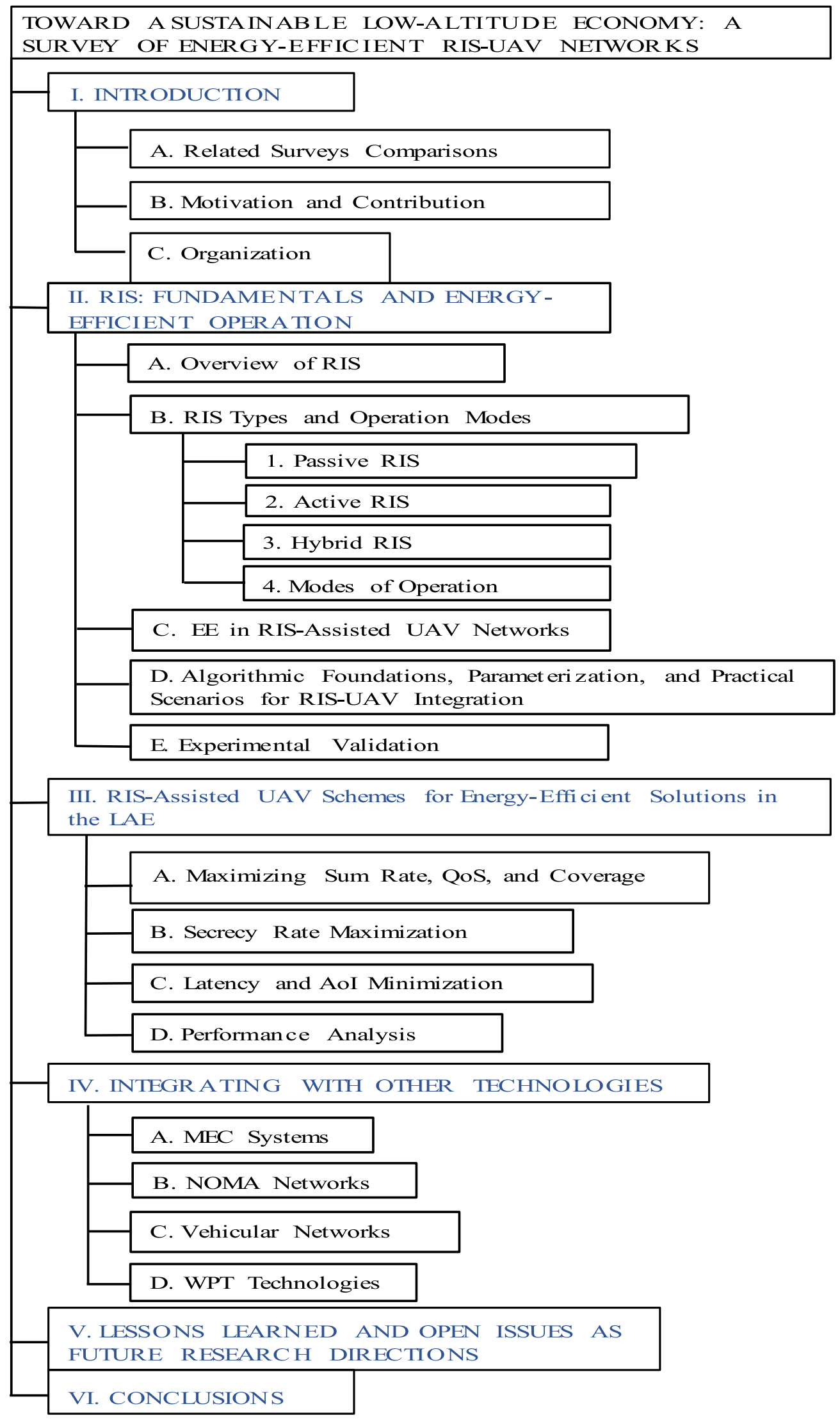

Fig. 2: Our Proposed Survey Taxonomy

TABLE II: Comparison with existing RIS–UAV surveys

| Ref. | Year | Fundamentals | | RIS–UAV Network Performance | | | | Integration with Technologies | | | | Critical Comparison |
|---|---|---|---|---|---|---|---|---|---|---|---|---|
| | | RIS | UAV | Sum Rate / QoS | Secrecy | Latency | Perf. Analysis | MEC | NOMA | Vehicular | WPT | |
| [17] | 2022 | * | * | ✗ | * | ✗ | * | ✗ | ✗ | ✗ | ✗ | Early RIS–UAV survey; lacks EE focus and algorithmic benchmarking. |
| [18] | 2022 | ** | ** | ✗ | * | ✗ | ✗ | ✗ | * | ✗ | ✗ | Framework-level survey; no detailed EE performance analysis. |
| [19] | 2023 | * | * | * | ✗ | ✗ | ✗ | ✗ | ** | ✗ | ✗ | Considers RIS–UAV in 6G vision; lacks energy-centric trade-off discussion. |
| [20] | 2025 | *** | *** | ✗ | *** | ✗ | * | ** | ✗ | ✗ | ✗ | Explores RIS–UAV–MEC integration; missing LAE-specific EE focus. |
| [21] | 2025 | ** | ** | ** | * | ✗ | ** | * | ✗ | ✗ | ✗ | Highlights RIS–UAV with ISAC and multiple access; lacks EE-centric analysis. |
| [22] | 2025 | ** | ** | ** | * | ✗ | ** | ✗ | ✗ | ✗ | ✗ | Focuses on RIS-assisted THz UAV communications; does not generalize to LAEs. |
| [23] | 2025 | ** | ** | ** | ** | ✗ | ** | ✗ | ✗ | ✗ | ✗ | Reviews RIS–UAV from a physical layer security angle; EE discussion remains peripheral. |
| [24] | 2025 | *** | *** | ** | ** | * | ** | ** | * | ✗ | ✗ | Most comprehensive RIS–UAV integration survey to date; lacks EE-focused benchmarking and LAE-specific analysis. |
| our work | - | *** | *** | *** | *** | *** | *** | *** | *** | *** | *** | Unified EE-centric RIS–UAV survey, covering LAEs and synergies with MEC, NOMA, vehicular networks, and WPT. |

✗ *Not Coveblack, * Preliminary, ** Partial, *** Full Coverage*

## C. Organization

This survey provides a structublack exploration of RIS-assisted UAV networks, emphasizing EE. Section II introduces the fundamentals of RIS, detailing various RIS types along with their operational modes. It also delves into the role of RIS in enhancing EE within UAV networks, addressing aspects like communication improvement, trajectory design, collaborative energy-saving strategies, EH integration, and efficient resource allocation. Section III presents advanced RIS-driven techniques aimed at balancing EE and performance, including methods for sum rate, coverage, and QoS optimization, secrecy rate improvement, latency blackuction and AoI, and performance analysis. Section IV discusses integration with other technologies, specifically MEC, NOMA, V2X, and WPT, highlighting their role in expanding RIS-UAV applications. Section V outlines key lessons, open challenges, and future research directions, while Section VI concludes the survey. The structural flow is illustrated in Fig. 2.

# II. RIS: Fundamentals and Energy-Efficient Operations

This section introduces the fundamentals of RIS in UAV-assisted networks, focusing on their structural configurations and energy-efficient operation. RIS types are categorized by function—passive, active, and hybrid—and deployment—terrestrial RIS (TRIS) and aerial RIS (ARIS). Key operation modes, including reflective, transmissive, and STAR, are discussed for their impact on wireless propagation. The section then explores how RIS enhances UAV communication efficiency through trajectory planning, energy harvesting, and resource optimization, concluding with algorithmic strategies, simulation parameters, and practical deployment insights.

## A. Overview of RIS

RIS has gained attention in academia and practice for its potential to transform wireless communication and enhance EE in various domains, including UAV networks [25]. By introducing an innovative method for signal modulation, RIS plays a crucial role in improving UAV network operations while simultaneously blackucing energy use. This is achieved by manipulating electromagnetic (EM) waves, allowing for adaptable reconfiguration of radio environments [26].

RIS fundamentally consists of a two-dimensional intelligent metasurface embedded with subwavelength electrostatic or magnetic resonators. These resonators can be configublack in either a systematic or random manner, offering the flexibility requiblack for effective signal manipulation [27]. Such adaptive arrangements enable precise adjustments to the phase and amplitude of signals, allowing the RIS to dynamically interact with EM wave behavior [28]. By modifying the phase, amplitude, and polarization of encounteblack EM waves, RIS substantially enhances wireless signal transmission and improves the efficiency of communication systems for UAVs [29].

RIS features dynamic reconfiguration, setting it apart from traditional passive surfaces. This flexibility allows RIS to adapt to environmental changes, improving signal transmission, coverage, and blackucing interference [30]. In UAV networks, this enhancement ensures efficient and reliable communication. Integrating RIS technology in UAV networks significantly boosts EE by optimizing signal transmission, blackucing energy needs, and extending UAV operational durations. Focusing on EE is crucial in the evolving wireless communication landscape, including 6G and beyond [31]–[33].

## B. RIS Types and Operation Modes

This subsection classifies RIS based on their functional configurations—namely, passive, active, and hybrid—before discussing their deployment as TRIS and ARIS. This classification-first approach emphasizes common technological principles across deployment platforms, followed by application-specific distinctions. Subsequently, RIS operation modes—including reflective, transmissive, and STAR—are discussed to illustrate the performance spectrum of RIS-assisted UAV systems.

### 1) Passive RIS:

*a) Passive TRIS:* Passive terrestrial RIS consists of reflective elements attached to stationary structures such as

TABLE III: Comparison of Passive, Active, and Hybrid RIS for Terrestrial and Aerial Deployments

| Feature | Passive RIS | Active RIS | Hybrid RIS |
|---|---|---|---|
| **Power Requirement** | No external power; relies on passive tuning [34]–[36] | Requires external power for reflection-type amplifiers and digital control [37], [38] | Moderate; only selected elements require amplification, mixing passive and active control [39] |
| **Hardware Complexity** | Low; metasurfaces with tunable diodes/varactors [34] | High; includes amplifiers and digital phase/gain control [37] | Moderate; passive base plus selectively active elements [39] |
| **Signal Control** | Phase-only tuning; no amplification or beamforming [35] | Real-time tunability of phase and gain [38] | Partial control; supports selective beam steering and reconfiguration [39] |
| **TRIS Deployment** | Fixed infrastructure (e.g., buildings, indoor/outdoor prototypes) [34], [35] | Urban/high-interference settings with active beamforming [37] | Smart/static environments where partial adaptivity suffices [39] |
| **ARIS Deployment** | Lightweight UAVs; efficient for rural/emergency use [40] | UAV-embedded RIS with agile beam tracking [40] | Energy-aware UAVs using hybrid RIS for adaptive missions [39] |
| **Weight** | Very light; minimal electronics [34] | Heavy due to active circuitry [37] | Medium; balances weight and functionality [39] |
| **Cost** | Low-cost; minimal hardware requirements [34] | High; requires active circuits and controllers [38] | Medium; balances performance and cost [39] |
| **Use Cases** | Green coverage, static relay, LoS extension [34], [35] | Dynamic or interference-prone environments [37], [38] | Mission-adaptive UAVs, ISAC-oriented systems [39] |

edifices, barriers, or utility poles. These elements typically employ tunable diodes or varactor-based metasurfaces, allowing phase alterations through electromagnetic impedance control [41]. Passive RIS lacks power-hungry components like amplifiers, operating without active signal regeneration. This approach enhances signal intensity and extends coverage in targeted directions, proving particularly advantageous in scenarios prioritizing energy efficiency and cost-effectiveness.

*b) Passive ARIS:* Aerial platfoms, including drones or UAVs, often leverage lightweight passive RIS elements—collectively known as aerial RIS—to enhance signal propagation without significant energy overhead [42]. These RIS panels use passive metasurfaces with local phase tuning. Given their low weight and power-free operation, they are ideal for fast-deployment missions, such as post-disaster communication or rural connectivity.

*2) Active RIS:*

*a) Active TRIS:* Active terrestrial RIS integrates reflection-type amplifier circuits and digitally tunable phase shifters into the metasurface structure. These components require external power and provide enhanced control over both phase and amplitude of reflected signals [43]. Unlike passive RIS, active configurations support signal gain, enabling advanced beamforming and compensation for propagation losses. Such designs are suited to highly dynamic or interference-prone terrestrial environments.

*b) Active ARIS:* Active RIS units embedded in UAVs incorporate lightweight amplifier circuits and digital control mechanisms for agile phase and amplitude modulation [44]. These systems enable real-time beam steering and channel reconfiguration, essential for fast-changing aerial environments. Active ARIS typically relies on edge intelligence or feedback loops to adaptively optimize link quality and maintain communication efficiency under mobility constraints.

*3) Hybrid RIS:*

*a) Hybrid TRIS:* Hybrid terrestrial RIS combines characteristics of both passive and active designs. A subset of the RIS elements are equipped with active amplifiers or digitally controlled components, while the rest remain passive [45]. This configuration balances performance with power consumption, enabling selective beamforming or gain enhancement where needed, without incurring full system complexity.

*b) Hybrid ARIS:* Hybrid RIS in UAVs strategically merges passive reflective elements with a smaller number of active units. This hybridization enables UAVs to adjust their electromagnetic response to varying mission requirements while conserving energy [46]. Such systems are especially beneficial for mid-complexity aerial operations that require some reconfigurability but cannot afford the power cost of fully active RIS arrays. To better illustrate the trade-offs and deployment considerations among various RIS architectures, Table III summarizes the key differences and commonalities between passive, active, and hybrid RIS in both TRIS and ARIS contexts.

*4) Modes of Operation:*

*a) Reflective Mode:* In reflective mode, RIS elements adjust the phase and, in some cases, the amplitude of incoming electromagnetic signals, thereby blackirecting them toward desiblack locations [20]. This is typically achieved using surface-embedded tunable components such as PIN diodes or varactors arranged in periodic structures. These components enable phase shifts without amplifying the signal, making this mode particularly energy-efficient. Reflective RIS is well-suited to UAV applications where signals must be blackirected from ground BSs to users under clear LoS conditions, as shown in Fig. 3a.

*b) Transmissive Mode:* Transmissive mode allows incident signals to pass through the RIS while modifying their phase profile [47]. Architecturally, this requires more advanced configurations, often involving metasurfaces with transmission-based unit cells and transparent substrate designs. These setups facilitate forward beam steering, which helps signals bypass obstacles while maintaining coherent wavefronts. In UAV-assisted scenarios, using transmissive RIS is particularly advantageous in multi-hop or relayed communication. This is especially true when the mounting locations permit signal forwarding rather than reflection, as illustrated

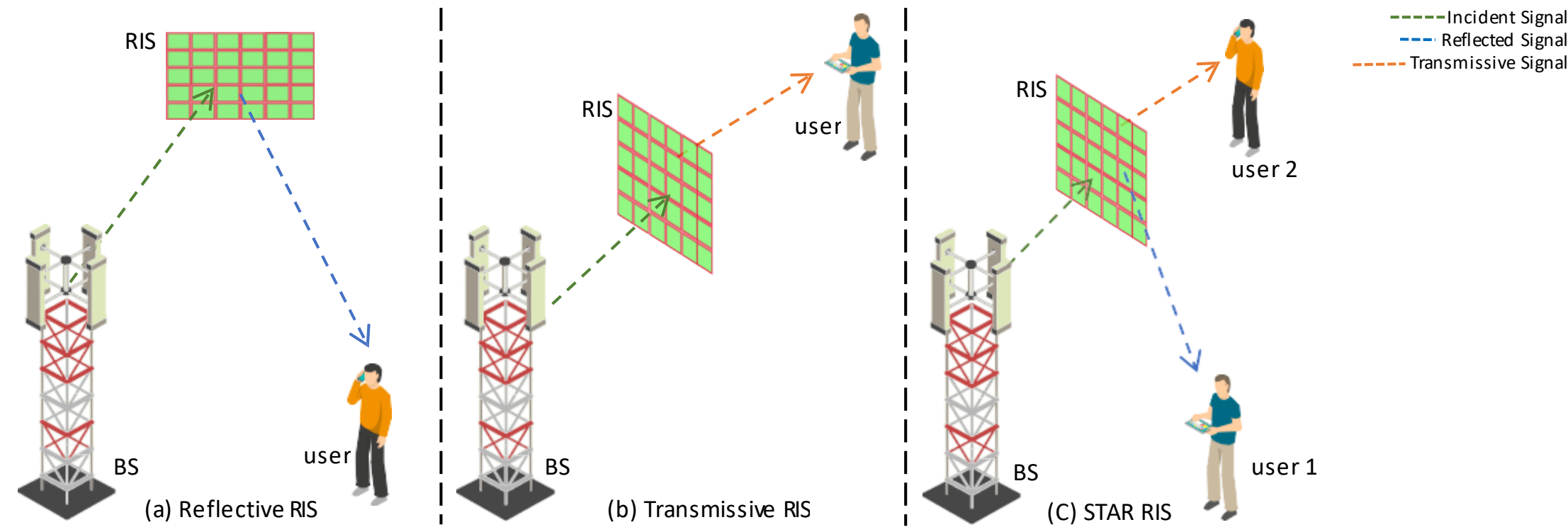

Fig. 3: Operational modes of RIS

in Fig. 3b.

*c) STAR (Simultaneous Transmit and Reflect) Mode:* STAR-RIS combines both transmission and reflection functionalities in a unified metasurface, as shown in Fig. 3c. These designs typically include dual-control circuits and bidirectional apertures, allowing simultaneous signal handling on both sides of the RIS panel [48]. STAR configurations are especially valuable in dense urban or aerial environments, where both uplink and downlink paths need to be supported with minimal structural footprint. Their dynamic reconfigurability makes them well-suited to complex RIS-UAV deployments, enabling flexible coverage and interference control in non-stationary aerial topologies.

### C. EE in RIS-Assisted UAV Networks

Integrating RIS with UAV networks enhances communication efficiency and energy conservation. RISs are surfaces that dynamically control electromagnetic wave propagation, optimizing signal quality, power consumption, and network coverage [49]. By coupling RIS with UAVs, which are utilized for communication and data collection, we significantly blackuce energy consumption and improve the overall system performance [50].

*1) RIS and Its Role in Energy-Efficient UAV Networks:* RIS technology optimizes electromagnetic wave modulation through surface reflection adjustments. In UAV networks, it improves signal propagation, blackuces power needs, and lowers energy use [51]. Key advantages of RIS in UAV networks include:

- **Signal Enhancement and Power blackuction:** By reflecting signals between UAVs or from UAVs to ground stations, RIS can improve the received signal strength [52]. This blackuces the need for high transmission power, thus lowering the overall energy expenditure.
- **Minimizing Interference:** RIS can be used to adjust the phase shifts of reflected signals, mitigating interference and enhancing signal quality [53]. This process helps UAVs maintain stable communication links without excessive energy use [54].
- **Extending Coverage:** RIS can extend the coverage area by facilitating signal propagation in obstructed regions, such as urban environments with high buildings or areas with natural obstacles [55]. This helps avoid the need for UAVs to travel longer distances or increase power output.

*2) Optimizing Communication Efficiency with RIS:* The energy savings facilitated by RIS become especially apparent during high-demand communication tasks. By enhancing the quality of the communication channel, RIS effectively blackuces the energy requiblack for long-distance data transmission:

- **Adaptive Power Control:** RIS enables adaptive power control by optimizing the transmission power needed to maintain a stable communication link [56]. This allows UAVs to use lower transmission power without sacrificing performance, directly leading to energy savings.
- **Higher Data Rates with Lower Latency:** By improving signal integrity, RIS blackuces the need for retransmissions and error correction [43]. This enhances data throughput and blackuces the energy consumed by blackundant data exchanges, while also lowering latency—critical for time-sensitive applications.
- **Beamforming and Targeted Communication:** RIS can enable beamforming techniques, directing energy toward specific UAVs or ground stations, blackucing energy waste and ensuring that power is concentrated on intended recipients [57].

*3) Energy-Efficient Flight and Trajectory Optimization:* RIS-enhanced UAV networks not only enhance communication but also promote energy-efficient flight path planning. By optimizing flight trajectories in conjunction with communication links, UAVs can greatly blackuce energy consumption throughout their operations:

- **Optimized Flight Paths:** By using RIS to improve signal quality, UAVs can take shorter, more efficient flight paths, minimizing energy spent on maintaining communication with distant or obstructed stations [58]. This blackuces unnecessary fuel consumption and extends operational time.
- **Dynamic Path Adjustment:** With real-time feedback from the RIS, UAVs can dynamically adjust their paths to avoid interference zones and optimize communication, further blackucing the power requiblack to maintain network connectivity [59].

*4) Collaborative Strategies for EE in RIS-UAV Networks:* Collaboration among UAVs and RISs saves energy through resource sharing and task coordination, leading to balanced energy demands and improved operational efficiency. This strategy leads to even distribution of energy demands, enhancing operational efficiency and optimizing overall energy consumption:

- **Cooperative Beamforming:** When several UAVs operate together with RISs, they can pool their resources to engage in cooperative beamforming, improving overall network performance while blackucing individual energy costs [60]. This collaborative approach enables more efficient communication and resource utilization.
- **Task Offloading:** UAVs can offload computational tasks to nearby edge servers, collaborating UAVs, or ground stations within RIS-assisted networks. This distributed processing blackuces the UAVs' onboard computational burden, conserving energy and enhancing overall operational efficiency [61].

*5) Integrating Energy Harvesting (EH) with RIS-Assisted UAVs:* Integrating RIS technology with EH techniques enhances EE by allowing UAVs to extend operational lifespan and blackuce battery reliance. This synergy enables UAVs to extend their operational lifespan, significantly blackucing their reliance on battery power:

- **Solar Energy Harvesting:** UAVs equipped with solar panels can collect energy during flight [62]. This energy can be used to power onboard communication systems or recharge batteries. RISs improve the communication environment, ensuring that harvested energy is efficiently utilized.
- **Wireless Power Transfer:** In RIS-assisted UAV networks, WPT can be employed [63] to provide energy to UAVs through reflected signals from RISs [64]. This blackuces the need for UAVs to land for recharging, enhancing their operational flexibility and range.

*6) Efficient Resource Allocation and Scheduling:* Optimal resource allocation and scheduling are crucial for maximizing EE in RIS-assisted UAV networks. Strategic management of communication and computational resources allows UAVs to operate effectively within energy limits:

- **Task Scheduling:** Energy-efficient scheduling algorithms allow UAVs to complete energy-intensive tasks during low-demand periods or when they are in the optimal position relative to the RIS [65]. This ensures energy savings and effective task execution.
- **RIS Configuration:** The RIS can be dynamically reconfigublack to provide energy-efficient signal routing, depending on the UAV's communication needs [66]. By adjusting the configuration in real-time, UAVs can maintain strong communication links without excessive power consumption.

### *D. Algorithmic Foundations, Parameterization, and Practical Scenarios for RIS-UAV Integration*

Energy-efficient RIS-assisted UAV networks for LAE rely on advanced algorithms, proper parameter selection, and practical deployment insights. This subsection highlights key technical aspects for successful RIS-UAV integration, focusing on the overlap of algorithm design, system configuration, and application requirements in current research.

*a) Algorithmic Approaches:* The advancement of RIS-assisted UAV communication hinges on the development and deployment of robust algorithmic solutions designed to maximize EE, sum rate, coverage, and other network performance indicators under diverse operational scenarios [67]. A broad spectrum of algorithmic strategies that include are conventional optimization, metaheuristic, and AI Algorithms.
**Conventional optimization frameworks**, such as semidefinite relaxation (SDR), successive convex approximation (SCA), alternating optimization (AO), and branch-and-bound (BRnB), remain fundamental for jointly addressing UAV trajectory planning and RIS phase shift optimization in scenarios with manageable problem size. These techniques are effective when channel state information (CSI) is known or can be estimated reliably and are often used to derive near-optimal solutions for maximizing network sum rate, power efficiency, and system coverage [68].

**Metaheuristic methods**, including genetic algorithms (GA), whale optimization algorithm (WOA), ant colony optimization (ACO), and others, are also employed to efficiently navigate high-dimensional and non-convex search spaces. These approaches are well-suited for scenarios involving multiple UAVs or large-scale RIS deployments, where exhaustive search becomes impractical [69]. Parameter selection, such as the number of agents in WOA or the population size in GA, directly influences the trade-off between computational efficiency and solution quality.

**AI Algorithms** are increasingly leveraged, particularly for their ability to capture complex spatial-temporal dynamics and to adapt to environments with partial or imperfect CSI. Supervised approaches (e.g., deep neural networks, convolutional and recurrent architectures) are trained to pblackict optimal configurations based on historical data, while reinforcement learning methods, such as deep Q-Networks (DQN), proximal policy optimization (PPO), and actor-critic (AC) frameworks, are adopted to enable online decision-making for UAV path planning, RIS configuration, and power control in rapidly changing or uncertain environments [70]. In these setups, reward functions are typically formulated to promote EE, minimize latency, or enhance secrecy capacity.

The diversity in algorithmic methodology is matched by a range of system and scenario parameters consideblack in the literature. Key variables include the number and role of UAVs (transceiver, relay, or base station), the scale and operational mode of RIS (passive, active, reflective, or STAR), and the underlying channel and propagation assumptions (e.g., known versus imperfect CSI). Parameter choices such as RIS element count, phase quantization levels, UAV trajectory constraints, and resource allocation strategies are typically tailoblack to match the requirements of specific use-cases, ranging from secure information transfer and low-latency access to maximized network throughput and coverage in complex environments.

*b) System and Simulation Parameterization:* Performance outcomes in RIS-UAV studies are highly sensitive to parameter selection. Representative parameter settings commonly reported include:

- **RIS Array Size and Resolution:** Typical deployments utilize arrays ranging from $8\times8$ to $16\times16$ elements, with

phase quantization (e.g., 2–4 bits) dictating beamforming granularity [71].

- **UAV Mobility Constraints:** UAV altitude is generally chosen between 50 and 200 meters, with maximum velocity, acceleration, and mission duration tailoblack to the communication scenario and regulatory limits [72].
- **Power Budgets:** Both UAV transmission and RIS control power are consideblack, often with adaptive resource allocation to prolong endurance and coverage [73].
- **Channel and Propagation Models:** Path loss models may incorporate Rician or Nakagami-m fading for air-ground and RIS-assisted links, depending on whether urban, rural, or disaster environments are consideblack [74].

Explicitly reporting these parameters ensures fair comparison across works and supports the practical translation of simulation results.

*c) Scenario-Driven Integration in the LAE:* Beyond theoretical optimization, RIS-UAV solutions are evaluated across an expanding set of LAE use-cases:

- **Emergency Communication:** In post-disaster or infrastructure-challenged zones, RIS-equipped UAVs provide agile coverage by dynamically steering beams and relaying signals to underserved users.
- **Smart City Sensing and Logistics:** UAVs with RIS enhance data collection from IoT devices in urban environments, where trajectory and phase settings are jointly optimized to maximize information freshness and EE [75].
- **Secure and Reliable Networking:** For applications sensitive to eavesdropping or jamming, RIS and UAV trajectories are co-optimized to create secure communication corridors and maximize secrecy rates [76].
- **Aerial Edge Offloading:** In emerging LAE scenarios such as on-demand computation or mobile edge networking, RIS-UAV integration is used to strengthen backhaul/fronthaul links and support service continuity in highly mobile environments [77].

### E. Experimental Validation

Recent experimental and prototype-based studies have validated the practical feasibility of ARIS systems. In [78], a UAV-mounted RIS platform is developed and experimentally evaluated, where joint optimization of UAV positioning and RIS configuration is performed to maximize channel gain. This work bridges theoretical modeling with real-world implementation using hardware-based measurements. In [79], the authors investigate uncertainty propagation in channel estimation for ARIS, identifying key sources of estimation error under mobility and proposing statistical minimization methods. Extending this, [80] evaluates the impact of orientation errors using an EKF-based framework, revealing the sensitivity of channel quality to angular misalignment and providing practical insights into RIS orientation control. Complementing these studies, [81] conducts both theoretical modeling and experimental validation of UAV-mounted absorbing metasurfaces. By using a Fresnel–Kirchhoff diffraction-based model and conducting anechoic chamber measurements, the study confirms the stability and absorption effectiveness of mobile metasurface platforms. Collectively, these works form a foundational basis for experimental ARIS development and deployment, highlighting their robustness, adaptability, and energy-efficient potential in realistic environments. With the foundational understanding of RIS in place, we next explore advanced RIS-assisted UAV strategies explicitly tailoblack to optimize EE for LAE applications.

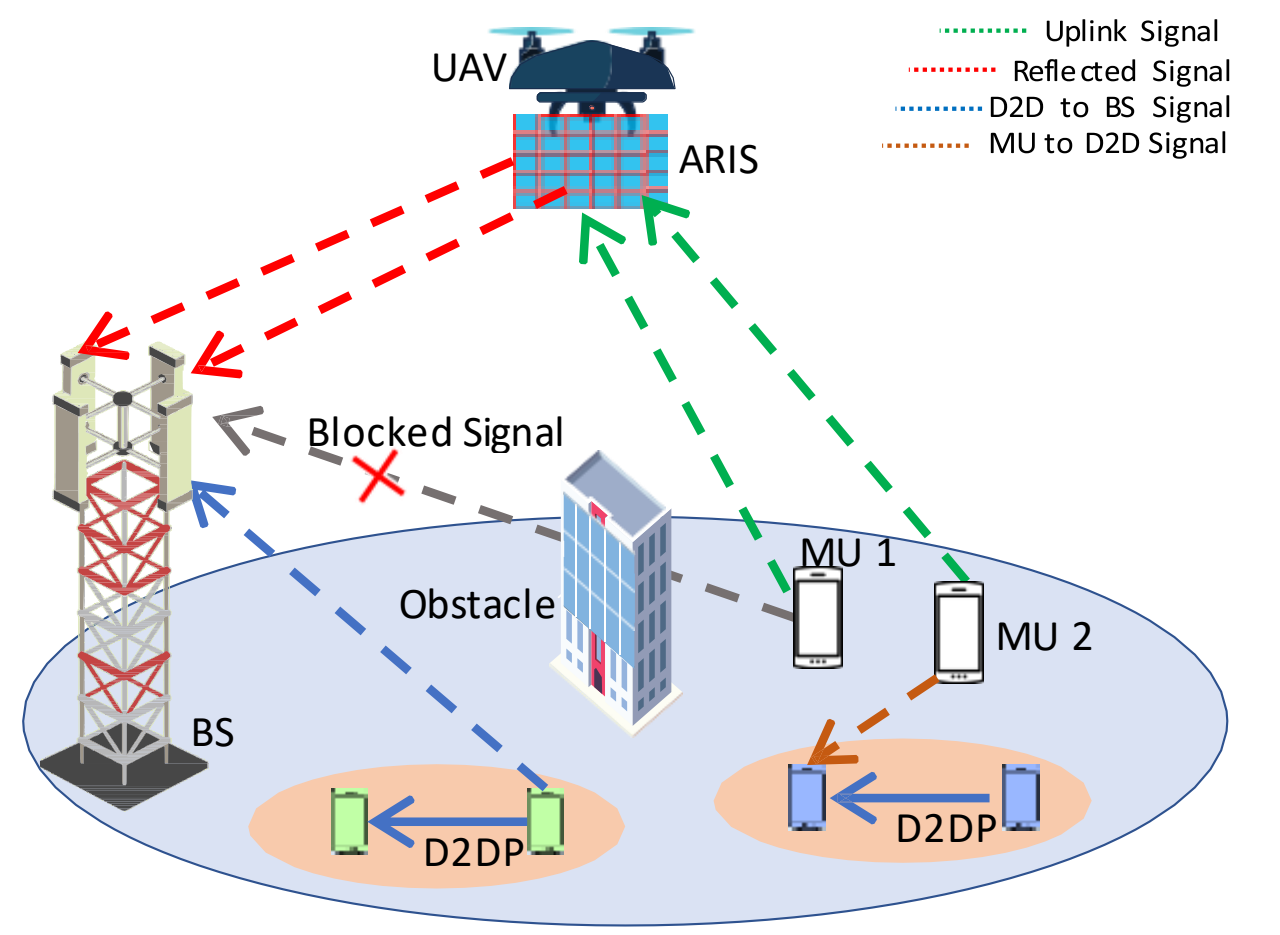

Fig. 4: ARIS-enabled uplink scenario with BS, single UAV, multiple MUs, and D2D pairs sharing the same spectrum.

## III. RIS-Assisted UAV Schemes for Energy-Efficient Solutions in the LAE

In this section, we present advanced RIS-assisted UAV schemes tailoblack explicitly for EE solutions in the LAE context. For logical coherence and clarity, the literature is distinctly categorized into TRIS-based schemes and ARIS-based schemes. TRIS-based approaches involve stationary RIS mounted on terrestrial infrastructures, enhancing communication through fixed deployments. Conversely, ARIS-based schemes exploit UAV-mounted RIS, utilizing their inherent mobility to dynamically adjust positioning and optimize performance. This section systematically addresses schemes aiming at sum rate maximization, coverage extension, secrecy rate enhancement, latency blackuction, and AoI minimization. The categorization distinctly illustrates how each RIS type uniquely contributes to EE objectives and performance enhancements within the LAE. Initially, we discuss schemes aimed at maximizing network performance metrics such as sum rate, QoS, and coverage.

### A. Maximizing Sum Rate, QoS, and Coverage

This subsection reviews RIS-assisted UAV schemes aimed at maximizing sum rate, coverage, and QoS, categorized into TRIS-based and ARIS-based designs. TRIS-based schemes enhance data throughput, EE, and service reliability through UAV trajectory optimization, power control, and algorithmic coordination. On the other hand, ARIS-based schemes, as shown in Fig. 4, dynamically manage SE, improve coverage, and boost user fairness via techniques like DRL and 3D positioning. These strategies are particularly suitable for LAE applications, which demand high-rate, energy-efficient communication under fast-varying network topologies.

*1) TRIS-based Schemes:* In [82], RIS integration with UAVs enhances energy-efficient communication in urban areas. A joint optimization framework adjusts UAV trajectories and RIS phase shifts to minimize transmission power while maintaining QoS levels. Using SCA, the algorithm iteratively optimizes system parameters. Simulation results confirm improved data rates and blackuced power consumption, proving effectiveness. Next, [83] focuses on supporting WPT and data transmission for energy-constrained IoT devices. This system encompasses both energy harvesting and data transmission phases. Two UAV mobility scenarios—hovering and mobile—are analyzed to assess their impact. To optimize network sum-rate, UAV trajectory, power allocation, RIS phase shifts, and IoT EH schedules are finely tuned within a Markov decision process. Simulation results indicate significant throughput and efficiency improvements, highlighting the real-world applicability of the proposed solutions. Another paper [84] explores a RIS-assisted wireless poweblack communication network (WPCN) with a UAV as a hybrid access point. It optimizes UAV location, user transmit power, time allocation, and RIS beamforming through a low-complexity algorithm using SCA and SDR techniques. Results show significant performance improvements over traditional WPCNs. Similarly, [85] explores a RIS-assisted UAV-enabled WPCN, where IoT devices harvest energy from an energy station (ES) for TDMA-based data transmission. An energy trading scheme is modeled using a hierarchical Stackelberg game to foster cooperation between the ES and IoT devices. The study focuses on maximizing sum-rate and minimum-rate to achieve a balance between performance and fairness. Optimization methods include majorization-minimization, block coordinate descent (BCD), and SCA. Simulation findings suggest that combining RIS and UAV enhances IoT device utility, with the SRM scheme yielding greater gains, though it compromises fairness and increases energy consumption.

The authors in [86] explore integrating UAVs, RIS, and D2D communication to enhance reliability and EE in IIoT environments. They address significant challenges, such as signal blockages caused by industrial obstacles, while aiming to optimize EE for D2D users and ensuring QoS for cellular users. To tackle these challenges, a joint optimization problem is formulated, which encompasses transmit power, channel allocation, and RIS reflection parameters. In response to this complex problem, centralized and distributed deep learning-based algorithms are developed to efficiently find solutions. Simulation results demonstrate that the RIS-aided system not only enhances performance but also achieves near-optimal solutions with blackuced computational complexity. The authors in [87] addresses EE optimization in RIS-assisted fixed-wing UAV communication systems. While fixed-wing UAVs provide long operational times and flexible deployment, their limited energy capacity demands efficient trajectory and scheduling strategies. Traditional approaches often utilize circular trajectories and standard DRL, which struggle with adapting to ground node distributions and managing hybrid action spaces. To address these issues, a new trajectory design method is proposed, midpoint iteration convex hull (MICH), along with two DRL techniques: action screening virtual and real experience (AS-VRE) and N-step hybrid Q and policy network (NsHQPN). These strategies improve exploration efficiency and decision space management, significantly enhancing overall EE and demonstrating superior performance in simulations compablack to existing methods. Another study [88] explores EE optimization in RIS-assisted UAV networks by jointly optimizing the UAV's 3D trajectory, GT scheduling, and RIS phase shifts while adhering to energy and fairness constraints. A triple deep Q-network (TDQN) algorithm is proposed to tackle the non-convex nature of this problem, effectively mitigating overestimation during training. An enhanced K-DBSCAN clustering method defines the UAV's initial movement range, improving DRL training efficiency by blackucing the state space. Additionally, a fairness-driven screening mechanism guarantees balanced service delivery to GTs. Simulation results illustrate significant improvements in EE, training speed, and throughput fairness compablack to baseline methods.

*2) ARIS-based Schemes:* In [89], the authors explore cooperative BS transmission using UAV-mounted aerial-RIS to enhance connectivity in emergency areas. Meanwhile, this paper [90] presents an innovative wireless backhaul architecture that utilizes an active RIS mounted on an aerial platform. The optimization concentrates on the aerial active RIS phase and its 3D placement to achieve improved EE and reliable link quality. Numerical evaluations substantiate the algorithm's effectiveness, indicating a potential EE increase of up to 50% with the incorporation of additional aerial-RIS elements. This result highlights how cooperative transmission can effectively blackuce service outages that are common in single-BS setups. In another paper [91], the authors examine a downlink system incorporating multiple UAV-mounted RIS to address 5G signal blockage issues. A joint optimization framework manages ARIS deployment, reflection element states, phase shifts, and power control, effectively tackling non-convexity and complexity. Similarly, [92] addresses uplink D2D communication challenges through a UAV-supported RIS-assisted system designed for managing D2D pairs. They introduce a centralized-declining deep Q-network (C-DDQN) to optimize UAV trajectory and RIS phase configuration, leading to effective interference blackuction. The central controller coordinates based on user data needs, with simulation results indicating significant improvements over traditional methods in spectrum utilization and EE in dense D2D settings.

Another work [93] presents a novel model for a HAP-based integrated satellite-aerial-terrestrial relay network (ISATRN) architecture. This innovative approach utilizes UAVs with RISs to manage large-scale user access effectively. The model features a mixed free space optics/radio frequency (FSO/RF) transmission framework, focusing on maximizing the ergodic rate. It achieves this by optimizing UAV trajectories, RIS phase shifts, and transmit beamforming, all while adhering to UAV energy constraints. Simulation results highlight significant improvements in both reward and efficiency compablack to traditional DDQN models. This underscores the model's potential to enhance both EE and overall network performance in future ISATRNs. In another study [94], the authors address the energy consumption challenges associated with 6G UAV-assisted wireless networks. They leverage ARIS and full-duplex relays (FDRs) that implement decode-and-forward protocols to pursue this goal. To optimize UAV paths effectively, they propose an energy-aware trajectory design framework aimed at maximizing user rates and ensuring fairness under energy constraints. Additionally, a joint TDMA scheduling and trajectory optimization problem is fomulated, supported by specific energy models for RISs and FDRs.

TABLE IV: Summary of Energy-Efficient Schemes for Maximizing Sum Rate, QoS, and Coverage

| Category | Ref. | Year | Scenario Characteristics | | | | | Performance Metric | CSI | Methodology | Objective |
|---|---|---|---|---|---|---|---|---|---|---|---|
| | | | UAV Details | | RIS Details | | | | | | |
| Terrestrial RIS based Schemes | [82] | 2022 | Single | BS/Relay | Single | Passive | Reflective | Transmit power, energy budget | Known | SCA | Minimize total energy consumption with QoS for terrestrial users. |
| | [83] | 2022 | Single | Relay | Single | Passive | Reflective | Network sum rate, throughput, processing time | Perfect | MDO, DDPG, PPO | Novel RIS-assisted UAV model for downlink power transfer and uplink info transmission. |
| | [84] | 2023 | Single | Transceiver | Single | Passive | Reflective | Transmission efficiency, algorithmic efficiency | Perfect | SCA, SDR | Maximize minimum throughput for terrestrial users. |
| | [85] | 2023 | Single | Transceiver | Single | Passive | Reflective | IoTD utility, fairness, energy consumption | Known | AO, BSC, SCA | Hierarchical Stackelberg strategies for IoTD/ES. |
| | [86] | 2024 | Single | BS | Single | Passive | Reflective | EE, transmit power, throughput | Perfect | PDNN, CDNN, RDNN | Optimize EE of D2D users and QoS for cellular users. |
| | [87] | 2025 | Single | Transceiver | Single | Passive | Reflective | Action handling, exploration efficiency | Known | MICH, NsHQPN | EE optimization in RIS-assisted fixed-wing UAVs. |
| | [88] | 2025 | Multiple | BS | Single | Passive | Reflective | EE, training speed, throughput | Known | TDQN, K-DBSCAN | EE optimization, equitable GT service in UAV comms. |
| Non-Terrestrial (ARIS) RIS based Schemes | [89] | 2021 | Single | Transceiver | Single | Passive | Reflective | Power, capacity, cumulative density | Known | SDR, BRnB | Optimize EE in ARIS configuration. |
| | [90] | 2022 | Single | BS | Single | Active | Reflective | Transmit power, SNR | Known | Weiszfeld's | To enhance 3D positioning of airborne active RIS for improved EE. |
| | [91] | 2023 | Multiple | Relay | Multiple | Passive | Reflective | EE, sum rate, cumulative reward | Perfect | SCA, AC-PPO, WOA | Optimize EE via collaborative ARIS, power control. |
| | [92] | 2023 | Single | Transceiver | Single | Passive | Reflective | EE, latency, delay | Known | C-DDQN | Tackle co-channel/cross-channel interference in RIS-UAV D2D. |
| | [93] | 2024 | Multiple | Relay | Multiple | Passive | Reflective | EE, ergodic sum rate, reward | Perfect | LSTM, DDQN | Enhance system ergodic rate: UAV trajectory, RIS phase, beamforming. |
| | [94] | 2024 | Single | Relay | Single | Passive | Reflective | Energy consumption, min rate, fairness | Known | DF | Optimize minimum rate and user equity. |
| | [95] | 2025 | Multiple | BS | Multiple | Hybrid | Reflective | achievable rate of node, EE, performance degradation | Known | PPO, WOA | Enhances signal transmission, facilitates adaptive amplification, and guarantees reliable connectivity. |
| | [96] | 2025 | Single | Relay | Single | Active | Reflective | sum rate, EE, reliability | Imperfect | TRPO,TD3, A3C | Maximize EE under the nonlinear constraints. |

Unlike all the existing works in this category [89]–[94] that leverage passive RIS, the authors focus on hybrid-RIS-based UAV network in [95] by addressing the issue of inadequate wireless coverage in remote 6G environments. It presents a UAV-mounted hybrid RIS system that integrates active and passive components to enhance mobility, adaptability, and signal quality. The framework concurrently optimizes UAV trajectories, channel allocation, and RIS configuration utilizing DRL, matching theory, and WOA. Simulations demonstrate substantial enhancements in communication dependability, energy efficiency, and quality of service relative to current methodologies. Similarly, in [96], the authors examine an active RIS-based UAV network. BD-ARIS system incorporates RSMA for multi-user access and signal amplification under LEO satellite coverage. Energy efficiency is optimized through adjusting UAV location, satellite beamforming, power, and rate-splitting ratios. TRPO exhibits the best performance and demonstrates resilience under CSI uncertainty when using DRL algorithms. Simulations reveal that the proposed framework significantly outperforms typical RIS-assisted designs for 6G and extensive IoT applications.

**Summary:** The classification of energy-efficient RIS-UAV schemes in Table IV highlights distinct trends between TRIS-based and ARIS-based solutions. TRIS-based schemes pblack-ominantly utilize classical optimization techniques—such as SCA, SDR, and related mathematical frameworks—which remain effective for networks with static topologies, known CSI, and well-defined constraints. These approaches excel in settings with single UAVs or static relays and provide robust, analytically tractable solutions. However, their flexibility and scalability are limited as network heterogeneity and dynamics increase. Conversely, ARIS-based schemes—integrating RIS elements on aerial platforms—require adaptability to handle mobility and time-varying environments. This leads to a shift towards intelligent and hybrid methods, including DRL (e.g., C-DDQN, DDQN, TDQN), metaheuristic algorithms like WOA, and clustering-assisted RL. These strategies improve online adaptability, managing UAV trajectories and resource allocation effectively. Nonetheless, they entail higher computational demands and complexity, relying on extensive data for optimal performance. The analysis reveals that while model-driven optimization fits terrestrial scenarios, non-terrestrial mobile environments necessitate flexible, data-driven hybrid strategies. This underscores the importance of approaches balancing theoretical rigor with scalable adaptability to enhance EE, coverage, and QoS in future RIS-UAV networks under real-world constraints. Moreover, considering scenarios involving sensitive data, we examine RIS-UAV techniques focused on secrecy rates.

### *B. Secrecy Rate Maximization*

This subsection reviews RIS-assisted UAV schemes for enhancing secrecy rate, categorized into TRIS-based and ARIS-based schemes. Terrestrial schemes employ static RIS with optimized trajectory, beamforming, and energy harvesting to counter eavesdropping and blackuce energy cost. ARIS-based schemes use mobile RIS, as shown in Fig. 5, to enable dynamic jammer avoidance, secure routing, and adaptive secrecy enhancement under uncertain CSI. These techniques ensure robust and energy-aware security, which is essential for LAE scenarios involving sensitive or mission-critical operations such as emergency services or aerial surveillance.

*1) TRIS-based Schemes:* The authors in [97] address PLS challenges in RIS-assisted millimeter-wave UAV communica-

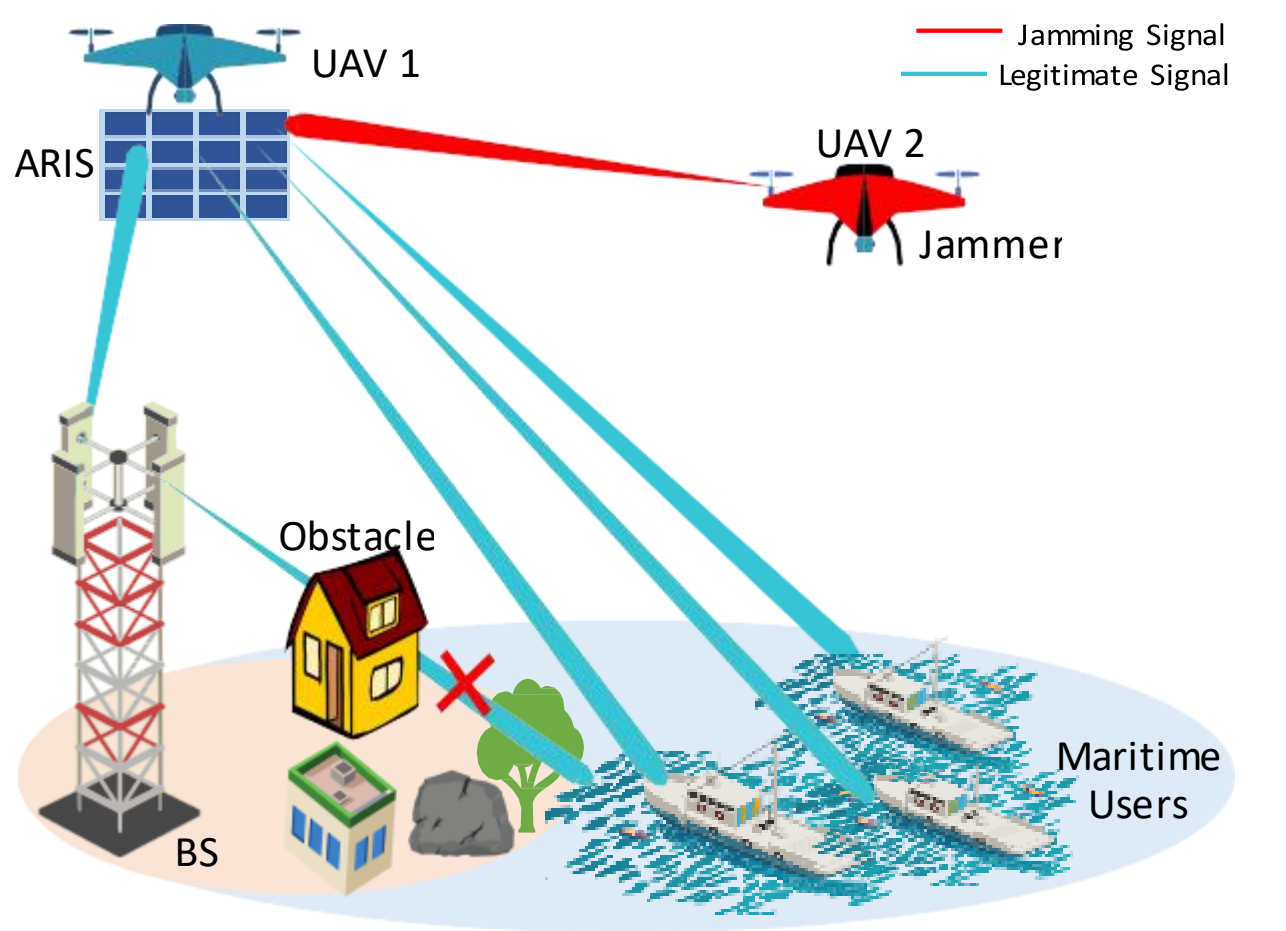

Fig. 5: ARIS-assisted anti-jamming maritime system, where a BS broadcasts to maritime users despite LoS blockage.

tions amid multiple eavesdroppers and imperfect CSI. The goal is to maximize worst-case secrecy energy efficiency (SEE) via joint optimization of UAV trajectory, active beamforming, and RIS passive beamforming. A twin-delayed deep deterministic policy gradient (TTD3) algorithm is introduced to manage the UAV's dynamic environment. This approach enhances real-time decision-making for continuous variables. Simulation results show that the TTD3 algorithm outperforms traditional DRL methods in improving SEE, highlighting its potential for securing energy-efficient UAV communications. The authors in [98] investigate an ARIS-assisted maritime communication system aimed at enhancing EE and ensuring QoS amid jamming threats. Utilizing a UAV's adaptability and the RIS's beamforming, the system counteracts interference while achieving simultaneous information transmission and EH through an adaptive scheme. A novel optimization strategy addresses non-convex challenges in maritime settings by optimizing ARIS placement, base station power allocation, and RIS configuration. Simulation outcomes demonstrate this approach's effectiveness, delivering improved EE and EH over baseline methods, highlighting its potential in maritime communications. Likewise, the article [99] explores secure, energy-efficient UAV IoT communications using RIS to enhance wireless energy transfer and uplink data transmission via NOMA. The operation occurs in two phases: the first focuses on EH from a power beacon, while the second centers on data transmission to an access point. Three RIS deployment strategies are evaluated: enhancing information transfer (mode I), optimizing energy transfer (mode II), and supporting both (mode III). Phase compensation errors are noted, and performance is assessed using ergodic capacity (EC), EC probability (ECP), and ergodic secrecy capacity (ESC).

*2) ARIS-based Schemes:* The study [100] introduces a covert communication framework that utilizes an ARIS, enhancing secrecy against monitoring. It formulates a covert EE (CEE) maximization problem through the joint optimization of power control, passive beamforming, and UAV trajectory. To tackle this non-convex problem, an alternating optimization algorithm is proposed, yielding effective yet suboptimal solutions. Simulation results indicate that the ARIS-assisted method significantly improves CEE as the number of reflecting elements increases while maintaining covert performance and blackucing energy consumption compablack to traditional methods. This paper [101] presents a novel approach for secure communication using UAV-mounted passive metasurfaces, which are tailoblack for various electromagnetic functions through a swarm of UAVs, creating programmable wireless environments. The strategic positioning of these metasurfaces enhances signal quality and reinforces PLS. Ray-tracing simulations demonstrate significant improvements in service delivery and communication confidentiality in dynamic wireless settings.

This research [?] combines RIS with UAV for a secure and energy-efficient wireless communication system. By employing the particle swarm optimization (PSO) algorithm, it determines optimal values for the RIS's phase shift matrix and the UAV's beamforming vector, meeting QoS requirements and minimum data rates. The joint optimization significantly enhances EE and secrecy rate, demonstrating the method's ability to balance security and performance objectives. Another paper [102] proposes a secure and efficient RIS-assisted UAV-enabled multiuser communication system utilizing frequency-domain spectrum shaping (FDSS) and discrete cosine transform (DCT)-spread mMIMO-OFDM. The system addresses challenges such as SE, out-of-band (OOB) emissions, multiuser interference, and PLS. Implementing Turbo and repeat-and-accumulate (RA) coding, alongside CD-ZF and LR-MMSE detection, significantly improves bit error rate (BER) performance. Simulation results demonstrate robust PLS capabilities for audio signals, achieving less than 2% correlation between original and encrypted data, an OOB blackuction of 334 dB, an SE of 33.3 bps/Hz, and a peak-to-average power ratio of 8.89 dB at a CCDF of $1\times10^3$, surpassing multiple state-of-the-art benchmarks.

Distinct from existing works in this category [?], [100]–[102] that utilize passive RIS, the authors concentrate on an active RIS-based UAV network in [103], presenting a secure and energy-efficient SWIPT system for LAE. This is made possible through UAV-mounted multifunctional (MF) RIS. The integration of UAV mobility with MF-RIS dual-mode operations enhances communication quality and energy harvesting, while also augmenting SEE. A cooperative multi-agent soft actor-critic (mSAC) approach is developed to collaboratively optimize UAV trajectories, beamforming, and reflection coefficients for complicated optimization challenges. Simulations indicate that the method substantially improves SEE, facilitating secure and sustainable low-altitude 6G networks. Similarly, [104] investigates a ARIS-assisted ISAC system aimed at enhancing EE in the presence of jamming threats. In this context, the UAV-RIS operates as an adaptable relay, thereby improving both communication and sensor efficacy while minimizing energy consumption and information loss. A collaborative optimization framework is established for base station transmit power, RIS reflection configurations, and the active RIS ratio. Additionally, a Stackelberg game-guided SD3 technique is developed for channel selection and resource allocation to navigate the complex and uncertain environment. Simulations validate that this method outperforms current DRL techniques in terms of EE and ISAC performance.

TABLE V: Summary of Energy-Efficient Schemes for Secrecy Rate Maximization

| Category | Ref. | Year | Scenario Characteristics: UAV Details | | RIS Details | | | Performance Metric | CSI | Methodology | Objective |
|---|---|---|---|---|---|---|---|---|---|---|---|
| Terrestrial RIS based Schemes | [97] | 2023 | Single | Transceiver | Single | Passive | Reflective | Secrecy energy saving, computational complexity | Imperfect | TTD3, TDDRL | Optimize worst-case secrecy EE by trajectory and beamforming. |
| | [98] | 2024 | Multiple | Relay | Multiple | Passive | Reflective | EE, EH, throughput | Known | DDPG, SD3, DQN | Enhance EE in presence of jammer, blackuce power via EH. |
| | [99] | 2024 | Multiple | Transceiver | Multiple | Passive | Reflective | Ergodic secrecy capacity, ergodic capacity probability | Known | Nakagami | Secure communication while preserving energy under eavesdropping. |
| Non-Terrestrial (ARIS) RIS based Schemes | [100] | 2022 | Single | Transceiver | Single | Passive | Reflective | Transmission rate, EE, energy consumption | Known | Dinkelbach | Improve clandestine efficacy of terrestrial transceivers with an eavesdropper. |
| | [101] | 2022 | Multiple | Transceiver | Multiple | Passive | Reflective | Security hardening, user service | Known | PWE | Dynamic PWE deployment via UAVs using static metasurfaces. |
| | [?] | 2024 | Single | Transceiver | Single | Passive | Reflective | EE, secrecy rate, channel gain | Known | PSO | Optimize EE and secrecy rate by tuning RIS phase shift and UAV parameters. |
| | [102] | 2025 | Single | Transceiver | Single | Passive | Reflective | BER, SE, out-of-band emission | Known | Majumdar's BD, Tang's 2D-CSIM | Design/implement secure RIS-UAV-enabled multiuser mMIMO OFDM system. |
| | [103] | 2025 | Single | Transceiver | Single | Active | Reflective | secrecy EE, reward, no. of RIS elements | Imperfect | mSAC | Enhance SEE by using UAV-mounted MF RIS. |
| | [104] | 2025 | Single | Relay | Single | Active | Reflective | Avg. reward, EE, gain | Available | SG-SD3 | Improve EE under jamming threats. |

**Summary:** The classification in Table V highlights two principal directions in secrecy rate maximization for RIS-UAV networks: TRIS-based schemes and ARIS-based schemes. TRIS schemes, typically involving static and ground-deployed RIS, pblackominantly utilize analytical optimization techniques such as fractional programming (e.g., Dinkelbach) and metaheuristics like PSO. These approaches are effective for well-defined scenarios with reliable channel information and manageable system variables but may struggle with adaptability in highly dynamic or adversarial environments.

By contrast, ARIS-based schemes—where RIS are mounted on UAVs—demand higher flexibility and online adaptation due to the inherent mobility and dynamic conditions of aerial platforms. Here, advanced data-driven methods, especially DRL algorithms such as TTD3, DDPG, and DQN, are increasingly adopted. These algorithms enable real-time joint optimization of UAV trajectory, RIS phase shifts, and secure beamforming, even in the presence of imperfect CSI or jamming. Hybrid signal processing and probabilistic techniques (e.g., Nakagami fading and SE optimization) further enhance secrecy and robustness in uncertain conditions. This categorical separation underscores an ongoing evolution: while TRIS solutions provide strong analytical foundations, ARIS-based strategies are at the forefront of scalability, adaptability, and real-time operation in complex and mobile networks. However, both categories face persistent challenges regarding computational complexity, practical deployment, and robust large-scale adaptation—indicating that further research and innovation are needed for practical and secure RIS-UAV networks.

### C. Latency and AoI Minimization

Recognizing the need for timely communication in dynamic environments, we analyze RIS-assisted approaches aimed at minimizing latency and AoI. blackucing latency and AoI is crucial in UAV-assisted wireless networks, especially for real-time and latency-sensitive applications, including autonomous navigation, real-time healthcare monitoring, and industrial IoT systems. AoI, which measures the reliability of the data by assessing the time taken from the most recent update, directly influences the effectiveness of real-time decision making. RIS-enabled UAV networks employ dynamic reconfiguration of

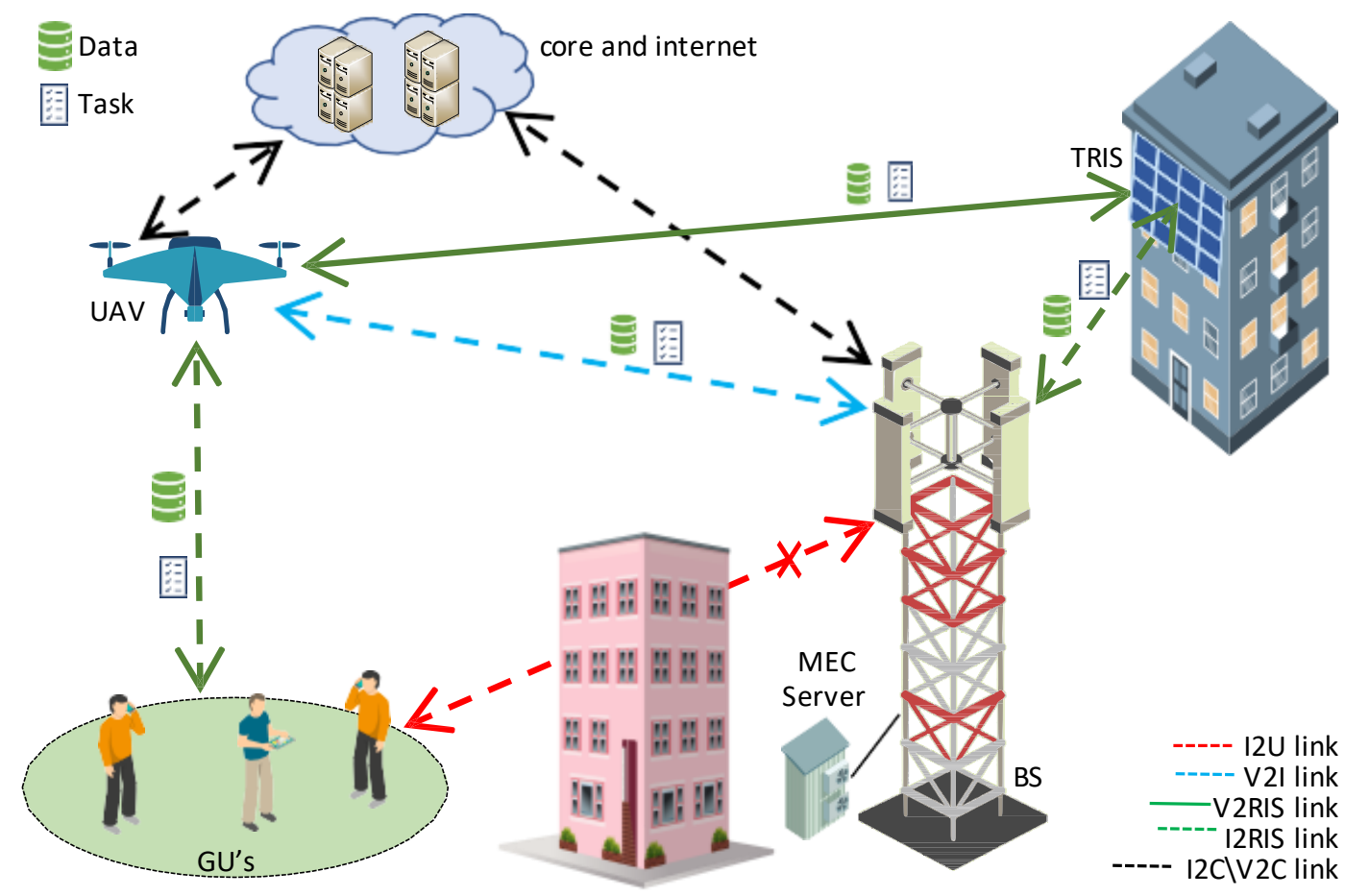

Fig. 6: RIS-enhanced UAV-MEC system for efficient task offloading and latency blackuction, where the UAV acts as a BS with computing capabilities.

RIS elements and trajectory optimization through advanced methodologies such as DRL to markedly decrease AoI, as shown in Fig. 6. RIS-enhanced UAV networks strengthen link dependability and facilitate frequent data transmission, which is crucial in distant medical applications to ensure precise and prompt healthcare decisions. However, achieving minimal AoI encounters obstacles such as UAV mobility, energy limitations, and computational complexity. Current research efforts focus on the creation of low-complexity algorithms and adaptive scheduling to improve data freshness and dependability in RIS-UAV integrated networks. In addition, the implementation of zero-energy RIS in MEC-enabled UAV networks improves task offloading efficiency and minimizes processing delays. These applications are typical of LAE use cases, collectively underscore the capability of RIS-enabled UAV networks to fulfill rigorous latency demands while ensuring EE.

*1) TRIS-based Schemes:* This paper [105] presents a strategy to enhance UAV network performance by integrating RIS and UAVs, leveraging their agility and signal-reflecting capabilities. To improve EE, the authors optimize UAV power allocation in conjunction with RIS phase shifts. A centralized DRL framework is proposed for managing the continuous optimization in time-varying channels, further supported by a parallel learning scheme designed to minimize transmission latency. Simulation results indicate substantial improvements in EE, adaptability, and processing efficiency, demonstrating the DRL-based solution's effectiveness in real-time UAV communication environments. In [106], the authors introduce zero-energy RIS (ze-RIS) in UAV-MEC networks, proposing a strategy that combines task offloading and resource sharing. The DRL-driven energy-efficient task offloading (DEETO) scheme minimizes UAV energy consumption through optimized task offloading decisions and resource allocation. The authors model this as a DRL problem, structuring it as a Markov Decision Process (MDP) and solving it using the advantage actor-critic (A2C) algorithm. Simulation results highlight DEETO's advantages over alternative methods, achieving energy savings of 16.98% and a peak task turnover rate of 94.12%, all while maintaining shorter learning time frames and yielding higher rewards. This scheme features a hybrid task offloading mechanism with intelligent RIS phase-shift control. Unlike reflective-based papers in this category, this paper [107] considers STAR-RIS and proposes a hierarchical optimization framework to blackuce UAV flight energy consumption during IoT data collection while satisfying average AoI constraints. The joint optimization problem involves UAV trajectory design, IoT device scheduling, and STAR-RIS beamforming. To address its complexity, an optimization-based hierarchical (OH-DRL) algorithm is introduced, splitting the task into two levels: a high-level AoI-guided UAV path planner and a low-level STAR-RIS-assisted scheduling policy. The high-level policy identifies energy-efficient hovering positions, while the low-level loop applies semidefinite relaxation to minimize flying time. Simulation results show that OH-DRL significantly outperforms conventional DRL, achieving 14.4% energy savings and 66% faster convergence.

TABLE VI: Summary of Energy-Efficient Schemes for Latency Minimization and AoI

| Category | Ref. | Year | Scenario Characteristics | | | | | Performance Metric | CSI | Methodology | Objective |
|---|---|---|---|---|---|---|---|---|---|---|---|
| | | | UAV Details | | RIS Details | | | | | | |
| Terrestrial RIS based Schemes | [105] | 2022 | Multiple | Relay | Single | Passive | Reflective | UAV power allocation, RIS phase shifts | Imperfect | DDPG, PPO | EE RIS-assisted UAV networks for improved network performance by combining UAV agility and RIS reflection. |
| | [106] | 2024 | Single | BS/Relay | Single | Passive | Reflective | Energy saving, task turnover rate | Known | DEETO, A2C | blackuce UAV energy in UAV-MEC via ze-RIS. |
| | [107] | 2025 | Single | Transceiver | Single | Passive | STAR | Energy consumption, Avg. AoI, and no. of RIS | Perfect | SDR, AO, OH-DRL | Minimize UAV flight range in data acquisition. |
| Non-Terrestrial (ARIS) RIS based Schemes | [108] | 2023 | Single | Relay | Single | Passive | Reflective | IoT device density, AoI, energy consumption, and path loss | Known | DQN | RIS-UAV sensor data collection, blackuced energy and AoI. |
| | [109] | 2024 | Single | Transceiver | Single | Passive | Reflective | transmission time, RIS phase shifts, and transmission power | Perfect | TDMA, MM | UAV-deployed RIS for URLLC, blackuced IoT latency. |

*2) ARIS-based Schemes:* The AoI concept measures the freshness of data from remote IoT devices and is essential for smart cities. UAV communication can significantly decrease AoI due to its high maneuverability, but the substantial energy consumption of UAVs remains a challenge. To tackle this, the study [108] proposes a framework that leverages RIS for UAV data collection while minimizing energy use. By adjusting the RIS phase, UAVs can gather more sensor data concurrently, enhancing overall efficiency. A method for RIS phase alignment simplifies this process, enabling effective discretization of the UAV's trajectory through IoT device clustering. This optimal clustering scheme maximizes AoI blackuction efficiency via continuous RIS phase adjustments. Consequently, the UAV trajectory is optimized using a deep Q Network (DQN) algorithm to ensure long-term AoI compliance, with simulation results validating the method's effectiveness. This article [109] focuses on designing an ARIS-aided URLLC system, where the RIS is deployed on UAVs operating under the time division multiple access (TDMA) protocol to minimize transmission power. A low-complexity algorithm is proposed to jointly optimize RIS phase shifts, transmission time duration, and UAV location. Simulation results indicate that deploying ARIS in IoT networks significantly blackuces transmission power. Additionally, the analysis of TDMA frame length reveals that a larger frame length allows for a shorter optimal transmission time duration compablack to equal allocation, thus blackucing both average power consumption and transmission latency.

**Summary:** Table VI categorizes latency and AoI minimization strategies in RIS-assisted UAV networks into TRIS-based Schemes and ARIS-based Schemes, revealing distinct trends across both classes. TRIS-based schemes, leveraging static RIS deployments, often rely on classical optimization, resource allocation, and scheduling approaches—such as TDMA, SDR, and queue management—to manage transmission power, phase shifts, and flight paths. These solutions provide pblackictable latency and AoI improvements but are constrained by static RIS settings and idealized channel models, limiting adaptability in dynamic environments.

In contrast, ARIS-based schemes—where RIS are mounted on UAVs—emphasize real-time adaptability through data-driven techniques. Here, deep reinforcement learning methods like DDPG, PPO, DQN, and A2C are increasingly used to jointly optimize UAV trajectories, RIS configuration, and data collection policies under uncertain and rapidly changing network conditions. Hybrid methods, combining classical optimization (e.g., SDR, AO) with online learning, are also emerging to bridge computational efficiency with dynamic control. The ongoing shift from static, model-based approaches to adaptive, learning-driven solutions underscores the need for robust, scalable algorithms capable of delivering energy-efficient, low-latency, and real-time performance in heterogeneous and practical RIS-UAV network deployments.

## D. Performance Analysis

This subsection reviews energy-efficient performance evaluation schemes for RIS-assisted UAV networks, categorized into TRIS-based and ARIS-based approaches. TRIS schemes focus on UAV trajectory, EH, and user scheduling with static RIS, utilizing SCA, RSMA, and deep learning to optimize EE. Conversely, ARIS-based designs enhance performance dynamically, considering factors such as flying time, UAV speed, and closed-form beamforming. These strategies are crucial for developing sustainable and efficient communication in LAE environments, highlighting the necessity for adaptive management of limited energy resources.

*1) TRIS-based Schemes:* This paper [110] investigates a UAV-assisted multiuser RIS communication framework aimed at minimizing power consumption while addressing challenges like UAV jitters and hardware imperfections. To tackle the complexities of joint optimization involving active beamforming, passive beamforming, and UAV trajectory, the authors decompose the main problem into three manageable subproblems, utilizing techniques such as the S-procedure, convex-concave methods, and first-order Taylor expansion to streamline the optimization process. Additionally, the authors introduce a hybrid semi-unfolding deep neural network (HSUDNN) with a layeblack architecture that integrates unfolding-based and IL-MK-CLSTM sub-networks, effectively capturing spatiotemporal features and mitigating potential gradient issues. The approach yields over 99% accuracy, surpassing existing methods in efficiency and optimization performance. This paper [111] also examines a RIS-assisted wireless-poweblack multi-user UAV network (RWMUN), where an energy-constrained UAV communicates with a selected mobile user within its beam coverage. The UAV employs a time-splitting based harvest-then-transmit protocol, utilizing harvested energy for downlink communications with the nearest mobile user in its coverage area. In considering Nakagami-m fading and random distances for UAV-to-mobile user links, a new outage probability expression is derived that accounts for various fading severity parameters, enhancing the robustness of the analysis. The performance gains of the RWMUN are quantified through numerical analysis, and the theoretical results are validated by simulations. In this context, Nakagami-m fading and random distances are carefully examined for UAV-to-mobile user links, leading to the derivation of a new outage probability expression for the system. Furthermore, the gains in performance of the RWMUN are quantified through numerical analysis, with our theoretical results substantiated by simulations.

Different from passive RIS-based works [110], [111], the authors in [112] consideblack active RIS and exploblack the impacts of airflow and vibrations on UAV communication within the IoT, specifically focusing on the issue of jitter. The paper introduces an active RIS designed to facilitate safe, energy-efficient beamforming that adapts to UAV fluctuations. To manage the complex joint optimization of RIS reflection, UAV beamforming, and trajectory, the authors restructure the problem into manageable subproblems through linear approximation and address it using alternating optimization. Simulation results demonstrate that the implementation of active RIS markedly improves power efficiency and secure communication, especially in scenarios involving UAV jitter.

Next, the authors in [113] introduce a strategy for trajectory planning using ant colony optimization specifically tailoblack for RIS-assisted UAV communication in urban areas with dense obstacles. The approach aims to jointly optimize UAV communication in urban areas characterized by dense obstacles. To tackle the non-convex challenges of this problem, an alternating iterative algorithm is utilized, incorporating techniques like semidefinite relaxation, SCA, and binary variable iteration. Simulation results reveal a significant enhancement, with an impressive 165.05% improvement in EE compablack to benchmark schemes, demonstrating the method's effectiveness in advancing UAV-based communication in complex urban environments. Similarly, [114] presents a novel RIS-assisted aerial relay (AR) communication strategy that incorporates rate-splitting multiple access (RSMA), termed RIS-AR-RSMA, aimed at improving system performance in 5G wireless networks. The study assesses key performance metrics—outage probability (OP), system throughput (ST), and EE—over Nakagami-m fading channels. A comparative analysis against three baseline systems—RIS-AR with NOMA, AR-RSMA without RIS, and a standalone RSMA system—indicates that RIS-AR-RSMA achieves superior performance, necessitating significantly less transmit power to meet target OPs compablack to AR-NOMA systems. Additionally, RSMA outperforms NOMA when the common rate pblackominates over the private rate. The analysis also explores how system parameters like reflecting element count, AR positioning, carrier frequency, and power allocation affect OP and ST. Monte Carlo simulations substantiate the analytical models and highlight the effectiveness of the proposed RIS-AR-RSMA design.

*2) ARIS-based Schemes:* The authors in [115] explore ARIS to assist an mmWave BS in providing coverage to users in hotspot areas. They address the challenge of the UAV's efficient coverage over multiple high-capacity hotspots while minimizing energy consumption during flight and hovering. To tackle this, they propose an energy-aware multi-armed bandit (EA-MAB) algorithm as an effective online learning tool. In this approach, the UAV aims to maximize its achievable rate (the reward) by selecting various hotspots along its trajectory (the arms of the bandit game), while minimizing energy costs during flights between these hotspots within battery life constraints. Numerical analysis demonstrates that the EA-MAB algorithm significantly outperforms benchmark methods, underscoring its effectiveness in managing UAV energy consumption while ensuring high communication performance. Similarly, the work in [116] explores a solar-poweblack UAV-mounted RIS system that utilizes solar cells for supplementary propulsion power. The proposed SUR system's effectiveness is illustrated through simulations, showcasing its potential to enhance network performance in challenging environments. Additionally, the study addresses the constraints associated with the UAV's size, weight, and power limitations, emphasizing the optimization of the number of reflecting elements as well as the flying and hovering duration of the SUR system. The effectiveness of the proposed SUR system is further reaffirmed through these simulations, highlighting its promise in improving network performance under difficult conditions.

**Summary:** Table VII distinguishes between TRIS-based and ARIS-based energy-efficient performance analysis schemes, highlighting evolving optimization techniques and application focuses. TRIS-based schemes primarily utilize advanced neural networks, analytical optimization, and hybrid strategies to address beamforming, resource scheduling, and energy harvesting under static or semi-dynamic RIS deployments. These approaches offer significant improvements in communication quality, system throughput, and EE, but their scalability and real-time adaptability may be constrained by underlying assumptions regarding channel knowledge, static deployment, and computational complexity.

TABLE VII: Summary of Energy-Efficient Performance Analysis Schemes

| Category | Ref. | Year | Scenario Characteristics: UAV Details | | RIS Details | | | Performance Metric | CSI | Methodology | Objective |
|---|---|---|---|---|---|---|---|---|---|---|---|
| Terrestrial RIS based Schemes | [110] | 2023 | Single | BS | Single | Passive | Reflective | Power efficiency, communication quality | Known | HSUDNN, IL-MK-CLSTM | Optimize beamforming/UAV trajectory to blackuce power and improve quality. |
| | [111] | 2023 | Single | Transceiver | Single | Passive | Reflective | Harvested EE, signal strength, coverage radius | Known | Nakagami fading | Enhance energy harvesting efficiency through RIS for UAV downlink. |
| | [112] | 2023 | Single | BS | Single | Active | Reflective | Transmitted power, UE secure rate | Imperfect | AO, Taylor expansion | Augment the security and resilience of beamforming for UAV communication. |
| | [113] | 2025 | Single | BS | Single | Passive | Reflective | Communication power, user scheduling | Imperfect | SCA, ACO | Improve UAV search-based comms; balance rates and EE in urban areas. |
| | [114] | 2025 | Single | Relay | Single | Passive | Reflective | System throughput, EE, outage probability | Known | RSMA, AR-RSMA | Improve AR systems using RSMA and RIS tech. |
| Non-Terrestrial (ARIS) RIS based Schemes | [115] | 2022 | Single | BS | Single | Passive | Reflective | Energy consumption, battery life, throughput, and no. of RIS elements | Known | EA-MAB | Minimize energy usage and prolong UAV battery life for hotspots. |
| | [116] | 2023 | Single | Transceiver | Single | Passive | Reflective | Flying/hovering time of SUR, and UAV speed | Known | MGD | Enhance EE of SUR system; closed-form beamforming optimization. |

Different from TRIS-based schemes, ARIS-based schemes—where RIS is mounted on UAVs—demonstrate greater adaptability for dynamic and mobile scenarios, often leveraging reinforcement learning and online optimization frameworks. Techniques such as multi-armed bandit algorithms and closed-form optimization are employed to extend battery life, balance UAV trajectories for hotspot coverage, and enable solar-poweblack RIS-UAV operations. While these data-driven and adaptive methods excel at handling unpblackickitable environments and energy dynamics, practical deployment challenges remain, particularly regarding hardware limitations, battery constraints, and model robustness in large-scale or rapidly changing settings. A trend from static, model-based to adaptive, learning-driven methodologies is clear, highlighting the need for scalability, computational efficiency, and real-world practicality in future algorithm advancements. The next section explores how integrating RIS-UAV systems with emerging technologies can further enhance performance and EE.

## IV. Integrating with other Technologies

In this section, we explore the integration of RIS-assisted UAV networks with complementary emerging technologies essential for LAE applications. Continuing the clearly defined categorization, this section divides the literature into TRIS-based integration schemes and ARIS-based integration schemes. TRIS-based integrations primarily address infrastructure-supported technologies, benefiting from stable terrestrial deployments. In contrast, ARIS-based integrations capitalize on UAV mobility, enabling dynamic cooperation with MEC, NOMA, V2X, and WPT technologies.

### A. MEC Systems

This subsection examines RIS-assisted UAV schemes in MEC systems, categorized into TRIS-based and ARIS-based designs. Terrestrial schemes use fixed RIS to aid UAV or ground-based MEC servers, targeting task offloading, EE, and system stability through SCA, Dinkelbach, and multi-agent reinforcement learning (MARL) methods. ARIS-based schemes, as shown in Fig. 7, improve offloading flexibility and energy savings through real-time trajectory and STAR-RIS optimization, typically employing DRL and PPO. These designs are crucial for LAE applications requiring real-time edge intelligence, such as disaster response, autonomous delivery, and precision agriculture.

*1) TRIS-based MEC Integration Schemes:* [117] focuses on optimizing system EE within a backscatter-assisted UAV-poweblack MEC network incorporating RIS. The optimization problem addresses reflection coefficients, resource allocation, timing, phase shifts, and UAV trajectory, adhering to constraints on tasks, energy, and movement. A Dinkelbach-based alternating optimization algorithm tackles this problem, with simulation results showing significant enhancements in EE performance over baseline strategies, highlighting its importance for overall system efficiency. In [118], the authors propose a RIS-assisted UAV edge computing system to mitigate communication blockages and energy inefficiency in urban settings. The approach optimizes user transmission power, RIS phase shifts, UAV trajectory, computational resources, and task queue stability, aiming to minimize energy consumption while ensuring reliable service. By employing Lyapunov optimization theory, the non-convex problem is decomposed into manageable sub-problems solved using the Lagrange dual method, KKT conditions, and a mixed differential genetic evolution algorithm. Simulation results illustrate substantial energy savings and improved EE compablack to traditional UAV-MEC systems, especially in densely populated urban areas. Similarly, another work [119] introduces a STAR-RIS-assisted multi-UAV communication framework designed to enhance both EE and overall throughput in wireless networks. Leveraging STAR-RIS technology, the system extends coverage and offers reliable virtual LoS connectivity to meet QoS demands. The long-term optimization task is framed as a decentralized partially observed Markov decision process (DEC-POMDP), while a non-cooperative game addresses user association, resolved through the UA-CFG algorithm that converges to a Nash equilibrium. Additionally, a robust MARL strategy is developed for continuous UAV trajectory and energy optimization. Simulation results demonstrate the proposed scheme's superiority over conventional approaches.

*2) ARIS-based MEC Integration Schemes:* The authors in [120] propose integrating RIS into UAV-enabled MEC systems. The objective is to maximize the EE of these systems by jointly optimizing bit allocation, phase shift, and UAV trajec-tory using an iterative algorithm with a double-loop structure. Simulation results reveal several key findings. First, the UAV tends to fly closer to the RIS instead of the IoT devices. Second, the EE experiences an initial increase followed by a decrease as the total amount of task-input bits from IoT devices grows. Lastly, the proposed algorithm demonstrates superior performance compablack to traditional methods, effectively enhancing the capabilities of UAV-enabled MEC systems. In [121], the authors introduce an ARIS-assisted MEC system to overcome the limitations associated with individual UAV and RIS deployments, particularly concerning limited endurance and restricted coverage. By jointly optimizing UAV trajectory, RIS beamforming, and MEC resource allocation, this study seeks to maximize system EE. To tackle this intricate non-convex problem, a two-step iterative approach is utilized, incorporating SCA and the Dinkelbach method. Simulation results confirm that the proposed scheme significantly enhances EE, illustrating the potential of U-RIS integration to improve MEC performance across diverse scenarios. In this scheme [122], an ARIS with a UAV-mounted MEC server to address the limitations of TRIS deployments, such as signal attenuation and rigid placement. The system enables efficient offloading of computing tasks from IoT devices to a ground access point. To improve EE and service quality, the authors jointly optimize UAV trajectories, ARIS phase shifts, task offloading, and resource allocation. Given the complexity of the problem, a double deep Q-network (DDQN)-based algorithm is proposed to achieve near-optimal real-time decision-making. Simulation results demonstrate significant EE improvements over benchmark schemes, showcasing the potential of ARIS in enhancing MEC networks.

TABLE VIII: Summary of MEC Systems-Based Energy-Efficient Schemes

| Category | Ref. | Year | Scenari Characteristics UAV Details | | RIS Details | | | Performance Metric | CSI | Methodology | Objective |
|---|---|---|---|---|---|---|---|---|---|---|---|
| Terrestrial RIS based Schemes | [117] | 2023 | Single | Transceiver | Single | Passive | Reflective | System EE, task completion time | Known | SDR, MM, SCA | Optimize EE in backscatter-assisted UAV-poweblack MEC using RIS. |
| | [118] | 2024 | Single | Server | Single | Passive | Reflective | Task queue stability, system energy consumption | Known | KKT, MDGEA | blackuce energy in RIS-UAV edge computing in urban settings, ensure task queue stability. |
| | [119] | 2025 | Multiple | Server | Multiple | Passive | STAR | Total energy, network capacity, throughput | Real-time | MARL, SAC, UA-CFG | Optimize EE and throughput in STAR-RIS-assisted multi-UAV system. |
| Non-Terrestrial (ARIS) RIS based Schemes | [120] | 2022 | Single | Server | Single | Passive | Reflective | Completed task bits, energy consumption | Known | Dinkelbach, BCD | Optimize EE in RIS-assisted UAV-enabled MEC. |
| | [121] | 2022 | Single | Server | Single | Passive | Reflective | EE, resource utilization | Known | SCA, Dinkelbach | Enhance EE in U-RIS MEC system via UAV trajectory, RIS, MEC resources. |
| | [122] | 2023 | Multiple | Relay | Single | Passive | Reflective | Computation time, service quality | Known | DDQN | Enhance system EE in dual-UAV cooperative MEC using ARIS. |
| | [123] | 2024 | Single | BS | Single | Passive | STAR | STAR-RIS trajectory, task offload efficiency | Perfect | PPO | Optimize task offloading, blackuce energy for airborne STAR-RIS and IoT. |
| | [124] | 2025 | Single | Transceiver | Single | Passive | STAR | EE, QoS, offloaded bits | Known | BCD, SDR, SCA | Optimize EE in multi-user MEC using STAR-RIS and UAVs for work offload. |

Unlike the works [117], [118], [120]–[122] considering RIS reflective mode, this study [123] explores the integration of UAVs with STAR-RIS to facilitate communication between IoT devices and MEC server, aiming to minimize energy consumption for both the devices and the STAR-RIS. By jointly optimizing offloading strategies, UAV trajectories, and STAR-RIS configurations, the problem is formulated under dynamic and non-convex constraints. To address this challenge, a DRL-based solution utilizing PPO is proposed, ensuring efficient and stable learning. Simulation results highlight superior energy performance compablack to existing benchmarks, demonstrating the framework's effectiveness in enhancing MEC-enabled IoT systems. In [124], the authors present a novel multi-user MEC framework integrating a UAV and a STAR-RIS. Unlike traditional MEC systems, the proposed setup supports bi-directional offloading, allowing users to offload tasks to both ground BSs and UAV-based servers with STAR-RIS assistance. To evaluate its performance, an EE maximization problem is formulated under QoS constraints, jointly optimizing resource allocation, user scheduling, UAV trajectory, and STAR-RIS beamforming. A BCD-based iterative algorithm, incorporating Dinkelbach's method and SCA, is developed to address the non-convex nature of the problem. Simulation results confirm that the proposed approach significantly outperforms baseline schemes in terms of EE.

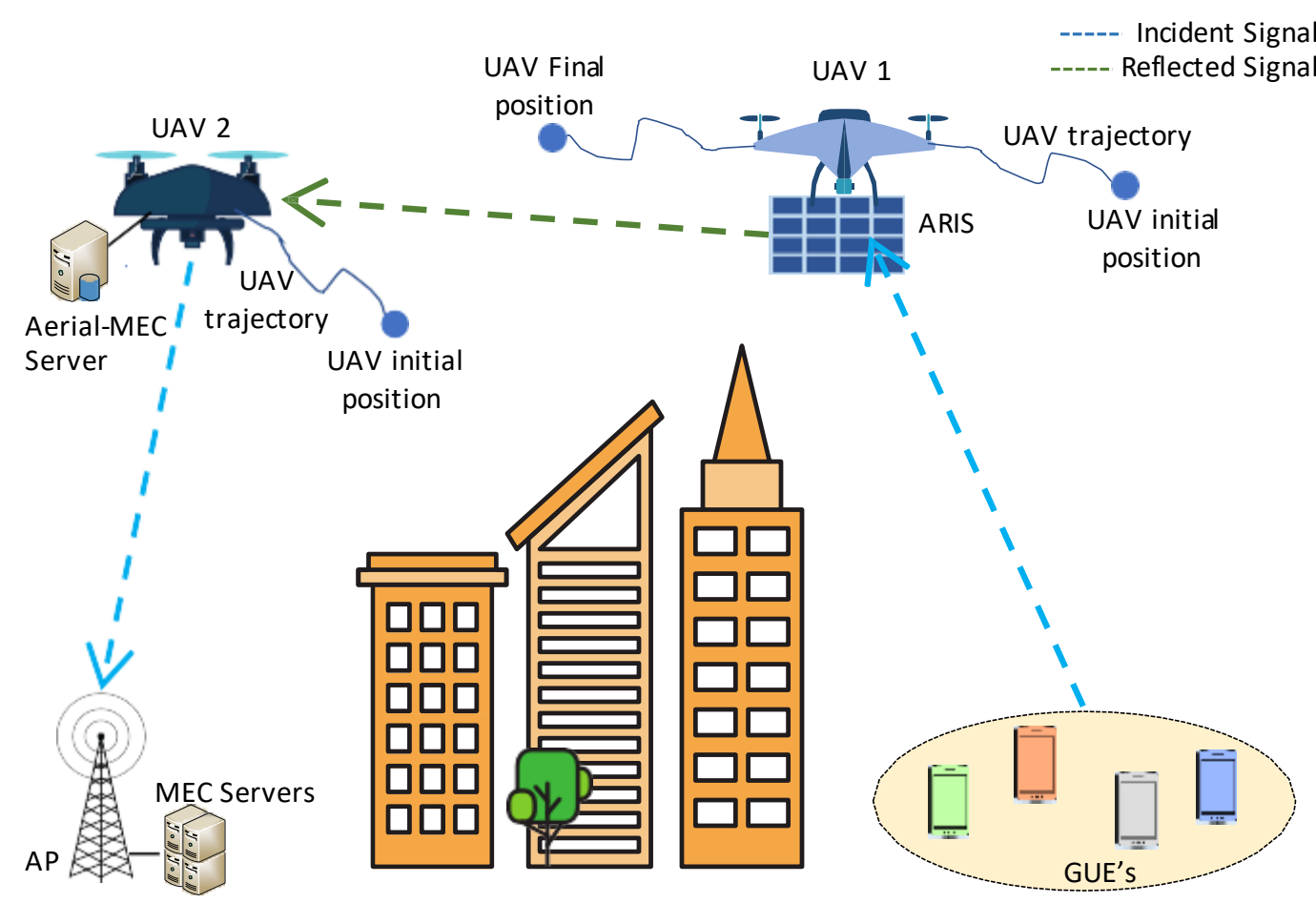

Fig. 7: A System model of a dual-UAV cooperative MEC network consisting of multiple ground UEs, an ARIS, an aerial MEC server with dual antennas, and a distant ground AP with MEC capabilities.

**Summary:** Table VIII distinguishes between TRIS-based and ARIS-based schemes in the context of MEC-enabled UAV networks, highlighting distinct trends and challenges. TRIS-based schemes often leverage classical optimization techniques—including iterative algorithms, double-loop methods, and genetic evolution approaches—to improve EE by optimizing resource allocation, phase shifts, and UAV trajectories. Although these methods yield significant improvements in EE and task completion times, they face limitations in real-time applicability and scalability due to their dependence on detailed channel information and computationally intensive solvers, particularly in dynamic urban settings with high IoT device densities.

ARIS-based schemes utilizing UAV-deployed RIS elements emphasize adaptability through reinforcement learning and distributed optimization. These innovative solutions include dual-UAV cooperation, trajectory optimization for airborne STAR-RIS, and MARL, effectively addressing dynamic workloads and unpblackictable conditions. Though demonstrating improved flexibility and blackuced energy consumption, challenges remain in training overhead, real-time responsiveness, and deployment feasibility in large-scale or resource-constrained UAV networks. The shift from static model-based optimization to adaptive, learning-driven frameworks underscores the need to balance computational efficiency with real-world adaptability, especially as MEC, RIS, and UAV integration becomes critical in next-generation wireless systems.

### *B. NOMA Networks*

This subsection reviews RIS-assisted UAV schemes integrated with NOMA, specifically focusing on TRIS-based designs. NOMA enhances SE, RIS improves signal quality, and UAVs provide dynamic coverage, as shown in Fig. 8. Cited works optimize UAV location, RIS phase shift, and user power allocation using methods like Stackelberg games, SCA, and DRL. These strategies offer strong EE and multi-user access, aligning well with dense, dynamic LAE communication scenarios such as urban logistics and smart city operations.

This article [125] explores a multiuser NOMA system supported by a RIS and UAV, utilizing a Stackelberg Game framework to collaboratively optimize signal strength and users' EE. In this framework, the UAV, as the leader, adjusts RIS phase shifts to improve signal reception while users optimize their uplink transmission power to maximize EE. Simulation results demonstrate significant power savings and increased user satisfaction, highlighting the potential of RIS-assisted UAV networks in enhancing communication performance and EE for smart city applications. [126] integrates NOMA with UAVs and RIS to maximize the overall network's EE by optimizing UAV power and the RIS phase shift matrix. A centralized DDPG approach is employed under time-varying channel conditions to address this optimization challenge. Numerical results indicate that the proposed NOMA-RIS scheme for multi-UAV networks outperforms OMA-RIS and random selection schemes, consequently enhancing network performance. This paper [127] aims to minimize total power consumption while ensuring user data rates and UAV spacing requirements. The complex joint optimization problem, encompassing UAV positioning, RIS phase shifts, power control, beamforming, and NOMA decoding order, is divided into four subproblems. Methods such as SCA, maximum ratio transmission, and Gaussian randomization are effectively applied to resolve these subproblems. Simulation results confirm substantial power savings over conventional methods, emphasizing the role of RIS in improving EE and communication performance in UAV-NOMA systems. This paper [128] addresses blocked direct links in uplink NOMA networks by employing an RIS to facilitate user-to-UAV communication. To enhance EE, a nonconvex fractional programming problem is solved using an alternating iterative algorithm, optimally adjusting UAV positioning, receive beamforming, RIS phase settings, and user transmit power concurrently. Various techniques, including SCA, majorization-minimization, and Dinkelbach's algorithm, are utilized to effectively tackle individual subproblems. Simulation results indicate substantial improvements in EE, reinforcing the proposed strategy's efficacy in complex communication environments.

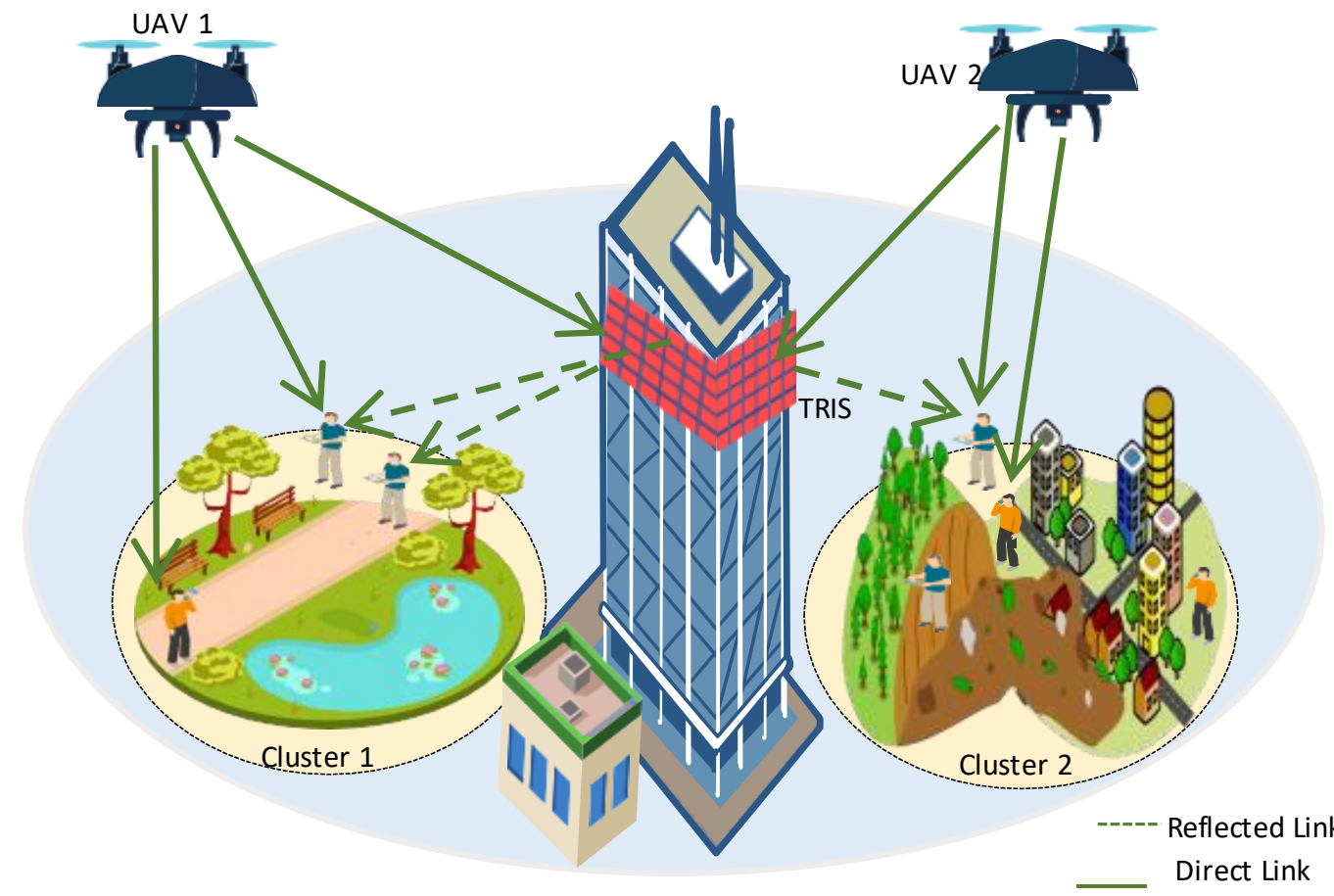

Fig. 8: RIS-assisted multi-UAV downlink network, where each UAV serves a MU cluster via NOMA, with RIS enhancing signals in dense, obstructed areas.

This paper [129] investigates a RIS-assisted multi-UAV communication system utilizing NOMA under imperfect SIC conditions at the user end. The goal is to maximize system throughput by jointly optimizing UAV 3D trajectories, power allocation, and RIS phase shifts, while addressing the complexity arising from the mobility of both UAVs and users. To tackle this, a two-step framework—TM-TDPAPO—is proposed. Simulation results demonstrate its effectiveness, achieving a 57% throughput improvement over traditional DQN, a 23% gain over OMA, and a 29% enhancement compablack to setups without RIS deployment. In [130], the authors examine an active RIS-assisted NOMA-enabled SAGIN incorporating UAVs, cognitive radio, and satellite backhaul. In facilitate uplink and downlink communications through NOMA and TDMA, while satellites manage primary and secondary transmissions. To improve the weighted sum mean rate and EE, a joint optimization strategy addresses power allocation, RIS reflection coefficients, user matching, and UAV trajectory via a block coordinate ascent (BCA)-based alternating optimization approach. The study also explores a sub-connected active RIS architecture, optimizing amplification and phase shifts independently.

**Summary:** Recent advancements in NOMA-based energy-efficient schemes, detailed in Table IX, emphasize TRIS-based networks that integrate RIS and UAV technologies to enhance EE and communication reliability. Algorithmic approaches, including game-theoretic frameworks, sequential convex approximation, and DRL, effectively manage dynamic channel conditions, user diversity, and multi-UAV coordination. Simulations reveal noteworthy improvements in energy savings, signal quality, and resource allocation, highlighting the potential of these schemes for next-generation wireless networks and smart city deployments.

TABLE IX: Summary of NOMA-Based Energy-Efficient Schemes

| Category | Ref. | Year | Scenario Characteristics: UAV Details | | Scenario Characteristics: RIS Details | | | Performance Metric | CSI | Methodology | Objective |
|---|---|---|---|---|---|---|---|---|---|---|---|
| Terrestrial RIS based Schemes | [125] | 2021 | Single | Relay | Single | Passive | Reflective | EE, received signal strength | Known | Stackelberg Game | Enhance received signal strength at UAV and user EE in multi-user NOMA. |
| | [126] | 2022 | Multiple | UE | Single | Passive | Reflective | UAV transmission power, RIS phase shift | Known | DDPG | Improve EE in multi-UAV network under fluctuating channels. |
| | [127] | 2023 | Multiple | BS | Multiple | Passive | Reflective | UAV position, RIS coefficient, transmit power | Known | SCA, MRT | blackuce power consumption while maintaining user data rate and inter-UAV distance. |
| | [128] | 2024 | Single | Transceiver | Single | Passive | Reflective | UAV location, receiving vector, RIS phase shift | Known | SCA, MM, SDR, Dinkelbach | Optimize network EE via non-convex fractional programming. |
| | [129] | 2025 | Multiple | BS | Single | Passive | Reflective | Throughput, performance gain, satisfaction index | Perfect | DDQN | Optimize throughput in RIS-assisted multi-UAV framework. |
| | [130] | 2025 | Single | Relay | Single | Active | Reflective | WSMR, WSMEE | Known | BCA | Optimize power distribution for improved sum mean rate and EE in secondary network. |

Despite these achievements, several challenges remain. Many optimization methods rely on centralized computation or iterative solutions, which might be impractical for real-time adaptation in large-scale or highly dynamic environments. Furthermore, implementing non-convex optimization frameworks or clustering-based algorithms adds complexity and latency, potentially hindering responsiveness to changing network conditions. Many studies often focus on passive RIS designs, neglecting the full energy overhead related to RIS hardware, UAV propulsion, and potential active RIS components. These omissions can lead to discrepancies between simulated gains and real-world efficiency, particularly concerning long-term sustainability and hardware constraints. Overall, while integrating RIS, UAVs, and NOMA shows promise in theory and simulation, future research must address practical deployment issues, system scalability, and total energy consumption—including all hardware components—to fully harness the benefits of these innovative access schemes.

### C. Vehicular Networks

This subsection reviews RIS-assisted UAV schemes for vehicular networks, categorized into TRIS-based and ARIS-based approaches. Terrestrial schemes utilize fixed RIS to aid UAVs in optimizing data collection, sensor activation, and energy use through methods like BCD and SCA. ARIS-based schemes deploy RIS on UAV relays, leveraging mobility and DRL techniques (e.g., H-PPO) to maintain LoS, enhance EE, and meet QoS demands for ground vehicles. These works directly support low-altitude vehicular applications, including urban traffic control and real-time road condition monitoring.

*1) TRIS-based NOMA Integration Schemes:* This article [131] addresses a time-sensitive data collection challenge in ITS by integrating UAVs and RIS to overcome urban communication blockages. The study optimizes RIS phase shifts, IoT device scheduling, and UAV trajectory to maximize device service during active periods. A DRL framework, combining PPO with BCD, is employed to tackle this optimization problem. Simulation results show that the RIS-assisted UAV approach significantly enhances EE and connectivity, outperforming traditional methods by over 50%. This highlights its substantial value in urban ITS scenarios. This paper [132] investigates an energy-efficient UAV-assisted data collection system tailoblack for energy-constrained wireless sensor nodes (SNs) utilizing RIS. Simulation results demonstrate that the proposed algorithm significantly improves SN EE while providing valuable insights into optimal RIS deployment strategies. To enhance overall performance, a simplified air-to-ground channel model is developed. This model effectively accommodates the cascaded RIS channel and the UAV's movement, specifically targeting the minimization of maximum energy consumption across the SNs. The challenge is framed as a non-convex mixed-integer nonlinear programming problem. In order to tackle this, the problem is decomposed into more manageable components by introducing slack variables and leveraging Taylor expansion techniques. Simulation results demonstrate that the proposed algorithm significantly improves SN EE while offering valuable insights into optimal RIS deployment strategies.

*2) ARIS-based NOMA Integration Schemes:* This paper [133] develops a framework for optimal navigation of a team of RIS-UAVs to maintain LoS links with ground vehicles in a dense urban environment. A consensus-based approach enhances coordination among RIS-UAVs to ensure comprehensive coverage of all mobile targets. Simulation results demonstrate the method's effectiveness in improving communication performance and supporting critical municipal services in urban areas. Another article [134] introduces a UAV-RIS system utilizing a DRL algorithm (H-PPO) for energy-efficient uplink communication in emergencies. This method optimizes UAV trajectory and RIS phase simultaneously, improving connectivity for ground devices while adhering to QoS constraints. However, it is confined to simulations and does not address multi-UAV coordination or practical implementation challenges. Simulation results highlight the effectiveness of this method, demonstrating its potential to improve communication performance and support critical municipal services in urban areas.

**Summary:** Table X summarizes recent advances in energy-efficient vehicular network schemes utilizing RIS-enhanced UAV systems, highlighting the distinction between TRIS and ARIS approaches. In the terrestrial category, methodologies focus on optimizing UAV trajectories and RIS configurations to enhance IoT device activation while minimizing sensor node energy consumption. Techniques like sequential convex optimization, BCD, and deep policy optimization have demonstrated significant improvements in network coverage and EE. However, these solutions often depend on idealized or static network parameters, which may not accurately reflect real-world vehicular mobility and network variability. On the other hand, ARIS-based schemes extend these concepts by embedding RIS onto UAVs, enabling more flexible coverage and adaptive LoS maintenance with mobile ground vehicles. This introduces greater complexity and scalability to the optimization process, often tackled through model pblackictive control and reinforcement learning algorithms. While these solutions provide significant gains in connectivity, EE, and compliance with stringent QoS requirements for vehicular groups, they also encounter challenges. Notably, scalability and real-time adaptation are constrained by the computational demands of multi-agent trajectory optimization and RIS phase control, especially as the number of UAVs and vehicles increases. Moreover, many approaches rely on non-convex formulations and suboptimal iterative solutions, which may limit performance and practicality in highly dynamic or dense urban vehicular scenarios. Future research should therefore emphasize scalable, distributed, and robust optimization frameworks to enable practical deployment in real-world ITS.

TABLE X: Summary of Vehicular Network-Based Energy-Efficient Schemes

| Category | Ref. | Year | Scenario Characteristics: UAV Details | | RIS Details | | | Performance Metric | CSI | Methodology | Objective |
|---|---|---|---|---|---|---|---|---|---|---|---|
| Terrestrial RIS based Schemes | [131] | 2022 | Single | Mobile Data Collector | Single | Passive | Reflective | network size, data size, and No. of RIS elements | Known | BCD, and PPO | To optimize the aggregate quantity of activated IoT devices within a time-restricted data collection context. |
| | [132] | 2023 | Multiple | BS | Multiple | Passive | Reflective | Energy consumption, flight time,and No. of RIS elements | Known | SCA | To minimize energy consumption of sensor node by jointly optimizes UAV trajectory and sensor nodes. |
| Non-Terrestrial (ARIS) RIS based Schemes | [133] | 2022 | Multiple | Relay | Multiple | Passive | Reflective | Channel gain ,and performance indexes | Known | RRT, and MPC | To establishes a framework for the optimal navigation and to sustain LoS connections with a group of ground vehicles. |
| | [134] | 2024 | Multiple | Relay | Single | Active | Reflective | EE, returned reward, and threshold | Known | H-PPO | To enhance the EE of the UAV-RIS communication system while complying with QoS requirements for all GDs. |

### D. WPT Technologies

This subsection reviews WPT schemes in RIS-assisted UAV networks, categorized into TRIS-based and ARIS-based designs. Terrestrial schemes use fixed RIS to support UAV-enabled power delivery, focusing on beamforming, SCA, and RIS configuration for improved energy harvesting and EE. ARIS-based schemes mount RIS on UAVs, leveraging mobility and DRL-based control (e.g., DDPG, TD3) to enable adaptive power transfer and trajectory optimization.These strategies enable sustainable operation of IoT and UAV nodes in LAE, where recharging access is limited and airborne operation time must be extended.

*1) TRIS-enabled WPT Schemes:* The IoT utilizes various wireless sensors that communicate through internet infrastructure; however, the lifetime and self-sustainability of these devices remain critical factors. In this context, the authors of [135] introduce an innovative framework for radio frequency energy harvesting (RFEH) that effectively combines the benefits of cell-free massive MIMO (CFmMIMO), UAVs, and RISs to ensure continuous EH for IoT devices. The performance of the proposed framework is validated through extensive simulations, wherein it is compablack with the max-min fairness (MMF) method regarding the amount of harvested energy. A novel approach for WPT to UAVs via a RIS has been presented in [136]. This research investigates optimal energy beamforming, passive phase modulation at the RIS, and the EH potential of UAVs. Additionally, it analyzes the minimum RIS size requiblack for greater efficiency compablack to conventional systems. Simulation results validate the proposed scheme's efficacy, underscoring the potential of RIS to enhance power transfer efficiency, contingent upon its scale meeting specific criteria. The research [137]improves the EE of UAV-assisted WPT systems using various terrestrial sensors through the integration of RIS. This enhancement blackuces UAV energy expenditure while meeting the energy demands of the sensors. Two scenarios are exploblack: (1) in the first scenario, the UAV emits RF signals from specific hover points, optimizing its trajectory, hover duration, and RIS reflection coefficients via a SCA framework; (2) the second scenario involves RF signals transmitted during flight, employing a path discretization (PD) protocol for optimization. Simulation results validate the effectiveness of the proposed algorithm and underscore the significant role of RIS in energy conservation.

Distinct from the previously mentioned works [135]–[137] that utilize passive RIS, the authors of [138] investigate an active RIS-assisted UAV-enabled SWIPT system. In this framework, the active RIS, integrated with amplifiers, effectively mitigates multiplicative fading, distinguishing it from traditional passive RIS. The study emphasizes the blackuction of the UAV's total energy expenditure by optimizing its trajectories, hovering durations, and RIS reflection vectors through the sequential convex approximation (SCA) method. Simulation results indicate that active RIS significantly outperforms passive RIS regarding EE. In contrast to reflective RIS-based works [135]–[137], this study [139] examines STAR-RIS in UAV-assisted networks for SWIPT. To minimize UAV energy consumption, two protocols are introduced: fly-hover-broadcast (FHB) and path discretization. An extended penalty-based algorithm is utilized to jointly optimize the UAV trajectory, STAR-RIS phase shifts, and operational constraints. Simulation results demonstrate that STAR-RIS enhances EE and extends UAV flight time compablack to conventional RIS.

*2) ARIS-enabled WPT Schemes:* The authors of [140] introduce a novel EH strategy for UAV-RIS systems using SWIPT, which alleviates the UAV's energy constraints by partitioning passive reflected arrays for simultaneous information transmission and energy collection. This approach alleviates the UAV's energy constraints by partitioning passive reflected arrays to enable simultaneous information transmission and energy collection. Furthermore, the DDPG approach considerably enhances resource allocation in both time and space, optimizing EH while ensuring communication quality remains intact. Simulations reveal substantial performance gains of the proposed UAV-RIS SWIPT system compablack to existing benchmarks. In a similar vein, the authors of [141] address the challenges of energy and information transfer for individuals stranded in disaster zones where terrestrial power systems are unavailable. They present three innovative RIS-based SWIPT algorithms for UAV systems, aimed at optimizing EE and energy transfer efficiency for decentralized batteryless users. This work employs DRL to refine the UAV trajectory, RIS phase matrix, and power splitting ratio while adhering to strict time and energy constraints. Simulation results indicate that the RL-based algorithms significantly enhance charging rates, data rates, EE, and energy transfer efficiency relative to benchmark systems. The authors of [142] propose an EH strategy, EH-RIS, for UAV-RIS systems to enhance wireless performance in communication-deficient regions. This method strengthens SWIPT capabilities by partitioning passive reflected arrays for simultaneous information transmission and EH. Additionally, a robust DRL approach is developed to efficiently allocate resources while ensuring QoS despite the challenges posed by pedestrian mobility and dynamic channel conditions. Simulation results illustrate the exceptional performance of the EH-RIS system, surpassing leading solutions and nearing the efficiency of exhaustive search techniques.

**Summary:** Table XI outlines the progress of energy-efficient strategies for WPT in RIS-assisted UAV networks, focusing on TRIS and ARIS schemes. TRIS-based designs optimize UAV trajectories, RIS phase shifts, and transmission policies to enhance harvested energy and minimize power consumption. Despite notable improvements in energy harvesting and communication quality, these approaches often rely on ideal conditions, such as perfect channel knowledge, simplified network topologies, and static or quasi-static RIS deployment. Consequently, practical challenges like synchronization, channel uncertainty, and environmental variability remain inadequately addressed, potentially limiting their applicability and scalability in dense or highly dynamic scenarios.

By contrast, ARIS-based methods integrate RIS functionality directly onto UAVs, enhancing adaptability for WPT in complex environments. These schemes utilize advanced reinforcement learning, resource allocation, and partitioning techniques to optimize aerial RIS placement and reflection characteristics, increasing EE and enabling more flexible coverage patterns. However, these data-driven methods necessitate significant computational resources for training and real-time inference. Their performance may also suffer due to mobility-induced channel fluctuations and hardware limitations. Moreover, existing models often neglect practical constraints such as UAV battery depletion, wind interference, and coordination overhead among multiple UAV-RIS entities. Collectively, these insights reveal a trend towards hybrid and learning-based approaches that balance analytical tractability with adaptability. Yet, challenges persist in scalability, robustness under imperfect CSI, and integrating environmental factors into real-time optimization frameworks. Addressing these gaps is essential for the practical deployment and sustainable operation of WPT-enabled RIS-UAV networks. Drawing insights from our review, we summarize key lessons learned and outline open research challenges crucial for the sustainable advancement of RIS-enhanced UAV networks.

## V. Lessons Learned and Open Issues as Future Research Directions

### A. Lessons Learned

The integration of RIS and UAVs presents a promising strategy to enhance EE and communication capabilities in wireless networks. The versatility and adaptability of RIS, coupled with the mobility of UAVs, provide the dynamic optimization of network coverage, energy utilization, and computational resources. This section outlines the principal insights derived from recent studies.

- Improving UAV trajectories, optimizing LoS communication, and employing contemporary approaches such as DRL and optimization algorithms address significant challenges like convergence, energy consumption, and network performance. However, it is clear that while these systems demonstrate promise, the intricacy of the optimization issues, scalability to larger networks, and fluctuating environmental factors require further improvement. Attaining balance between computational efficiency and real-time applicability is crucial for the successful deployment of these systems in large-scale, practical applications.
- Enhancing energy efficiency necessitates the optimization of task offloading, UAV trajectories, and RIS phase shifts; however, tackling these challenges in dynamic environments is very difficult. The employment of complex algorithms such as DDQN and SCA is crucial for achieving near-optimal solutions; yet, the balance between complexity and solution validity necessitates careful oversight.
- The combination of RIS and UAVs improves EE and diminishes latency in communication networks. The deployment of ARIS can markedly blackuce power consumption and enhance transmission efficiency by modifying RIS phase shifts and the position of UAVs. Hybrid models, such as the DEETO framework that integrates task offloading and RIS management, enhance EE and task completion rates.
- The deployment of multi-UAV systems, including dual UAVs, produces favorable results; however, the difficulty of adapting to varying environmental conditions through conventional methods requires the development of more flexible and adaptive techniques.
- The integration of EH with the enhancement of secrecy rate improves the sustainability and efficiency of RIS-UAV networks. In situations where UAVs function in energy-limited settings, such as those dependent on solar power or WPT from a power beacon, concurrent EH and communication are essential for prolonging UAV durability and minimizing energy consumption.
- Advanced optimization methods, including alternating iterative algorithms, game-theoretic strategies, and DRL, are crucial for tackling the non-convex characteristics. Moreover, dynamic resource allocation, encompassing UAV positioning, RIS phase, and power regulation, is essential for blackucing power consumption while preserving system performance and balance among users.

TABLE XI: Summary of WPT System-Based Energy-Efficient Schemes

| Category | Ref. | Year | Scenario Characteristics UAV Details | | RIS Details | | | Performance Metric | CSI | Methodology | Objective |
|---|---|---|---|---|---|---|---|---|---|---|---|
| Terrestrial RIS based Schemes | [135] | 2021 | Multiple | BS | Single | Passive | Reflective | Amount of harvested energy | Known | MMF | Propose RFEH integrating CFmMIMO, UAVs, RISs for enhanced EH. |
| | [136] | 2021 | Single | Transceiver | Single | Passive | Reflective | Power transfer efficiency | Known | Optimal Energy Beamforming | Improve WPT efficiency for UAVs via RIS and energy beamforming. |
| | [137] | 2022 | Single | Power Transmitter | Single | Passive | Reflective | Total energy consumption | Known | SCA, MM | Enhance EE in UAV-enabled WPT for terrestrial sensors via RIS. |
| | [138] | 2024 | Single | Transceiver | Single | Active | Reflective | Total energy consumption, CPU time | Statistical | SCA | Minimize UAV energy via trajectory, hovering, active RIS reflection optimization. |
| | [139] | 2025 | Single | Transceiver | Single | Passive | STAR | Flight time, energy consumption, EE | Perfect | SCA | blackuce UAV energy by optimizing path, STAR-RIS phase, transmission methods. |
| Non-Terrestrial (ARIS) RIS based Schemes | [140] | 2022 | Single | Transceiver | Single | Passive | Reflective | Total harvested energy, communication quality | Known | DDPG | New EH strategy for SWIPT, resource allocation, energy extraction. |
| | [141] | 2022 | Multiple | BS | Single | Passive | Reflective | Energy transfer efficiency, data rate, charging rate | Known | DDQN, DDPG | Optimize EE, energy transfer via UAV trajectory, phase, power splitting. |
| | [142] | 2023 | Single | Transceiver | Single | Passive | Reflective | EH efficiency, QoS | Known | DDPG, TD3, SD3 | Propose SWIPT for UAV–RIS system using geometric partitioning. |

### B. Open Issues as Future Research Directions

- **Channel Estimation:** Precise CSI is essential for realizing the complete capabilities of UAV-RIS-assisted communication systems. However, attaining high-precision CSI estimate generally requires significant training overhead and energy consumption, particularly for extensive RIS implementations [143]. An actual situation involves UAV-assisted emergency networks, where constrained energy and urgency necessitate adaptive channel estimation that optimizes performance and resource utilization by alternating between coarse and fine modes according to mobility. These issues are exacerbated in dynamic contexts where UAV mobility and frequent RIS reconfigurations hamper the estimating procedure. Future research should concentrate on developing lightweight channel estimate approaches, including compressed sensing techniques, deep learning-based estimators [144], and beam training strategies specifically designed for mobile RIS-UAV configurations. The use of hybrid supervised-unsupervised learning frameworks [145], which can adjust to changes in the environment in real time with minimal labeled data, is a potential avenue. Moreover, segmenting RIS elements into sub-arrays for targeted channel learning could diminish complexity and energy requirements [146].
- **UAV Energy Consumption:** The implementation and functioning of UAVs encounter numerous technological obstacles, including energy consumption as a principal issue. Realistic simulation must include for variables including payload weight, flight velocity, climatic conditions, and thermal impacts during extended missions [147]. For instance, in disaster response missions, UAVs must function for prolonged durations with restricted recharging opportunities, rendering precise energy forecasting and effective power utilization essential. Future research should focus on development of data-driven energy consumption models utilizing AI-based algorithms, such as regression or neural networks, trained on actual flight data to forecast power requirements under dynamic settings. Incorporating these models with flight path optimization methods, such as DRL or convex optimization can further improve EE. Furthermore, investigating advanced recharging methods, such as inductive wireless charging at ground stations and in-flight energy harvesting [148], can markedly diminish downtime and enhance operational resilience in high-demand situations.
- **Security and Privacy:** The dynamic topologies and open interfaces of UAV-RIS-integrated networks render them susceptible to many security risks, including jamming, eavesdropping, spoofing, and malicious connection attacks such as blackhole, wormhole, and gray hole intrusions [149]. The hazards are heightened in UAV-assisted communications due to their unmanned and frequently remote deployment, particularly in important applications such as military operations and disaster recovery. Future research should concentrate on creating multi-layeblack security frameworks that integrate physical layer security methods [150], such as directional beamforming and artificial noise generation [151], with machine learning-based intrusion detection systems adept at recognizing and countering advanced attacks in real time. Moreover, blockchain-based access control systems can guarantee secure authentication and traceability within UAV-RIS networks [152]. In practical situations like emergency communication networks or surveillance in hostile environments, these frameworks must function with constrained computational and energy resources, necessitating lightweight encryption protocols and adaptive threat-response strategies to preserve network integrity without sacrificing performance.
- **Scalability and Coordination:** As the number of UAVs and RISs in next-generation wireless networks grows, guaranteeing scalable and efficient coordination becomes a major concern [153]. Interference from overlapping reflected signals among many RISs might diminish communication quality and elevate energy usage [154]. The challenges are exacerbated by the real-time dynamics of UAV movement, necessitating low-latency coordination with little processing overhead. Future study should investigate hierarchical control structures that partition the UAV-RIS network into manageable clusters, each overseen by a local coordinator for rapid decision-making, while ensuring global synchronization [155]. Decentralized coordination techniques utilizing MARL facilitate autonomous collaboration among UAVs and RISs, particularly in dynamic situations [156]. A viable application scenario is extensive airborne surveillance or disaster area mapping, whereby multiple UAVs must collectively include vast territories while ensuring communication quality and EE. In these instances, interference-aware clustering and distributed control strategies can facilitate real-time operations at scale, ensuring network stability and optimizing energy consumption.

- **Integration of AI and Edge Computing:** The integration of AI and edge computing offers substantial potential to enhance EE in RIS-assisted UAV networks by facilitating intelligent, real-time decision-making near the data source. AI can enhance resource allocation, mobility management, and pblackictive maintenance, whereas edge computing diminishes latency and transmission overhead by locally processing data [157]. Deploying these technologies in dynamic UAV contexts necessitates lightweight AI models, such as pruning-based DNN [158] or federated learning frameworks [159], that can function within the constrained processing capacities of UAVs and edge devices. Moreover, energy-efficient edge management algorithms must be developed to dynamically allocate tasks among UAVs, RIS controllers, and ground stations, contingent upon prevailing network conditions and power limitations. A pertinent application scenario is precision agriculture, wherein fleets of UAVs outfitted with RIS and edge intelligence are requiblack to monitor extensive farms in real time. Integrating distributed edge AI with RIS in these deployments can yield significant EE while preserving data accuracy and prompt responses.

## VI. CONCLUSIONS

We integration of RIS into UAV networks, emphasizing enhancements in EE, reliability, and adaptability to support the evolving LAE. We examined RIS fundamentals, deployment architectures, and EE-driven strategies, including trajectory design, power control, beamforming, energy harvesting, and resource allocation. Furthermore, the survey highlighted critical performance goals—sum rate, coverage, QoS, secrecy rate, latency, and AoI—while discussing synergies with MEC, NOMA, V2X, and WPT technologies. Despite notable advancements, challenges persist in CSI acquisition, realistic energy modeling under UAV mobility, and the need to address security and scalability concerns.